\documentclass[%
 reprint,
 superscriptaddress,
 floatfix,
 amsmath, amssymb,
 aps,
 prx,
]{revtex4-2}
\usepackage{enumerate, physics}
\usepackage[justification=justified,
   format=plain]{caption} 
\usepackage{graphicx, subcaption, ragged2e}
\usepackage{dcolumn}
\usepackage{bm}
\usepackage[colorlinks=true,linkcolor=blue,citecolor=blue,urlcolor=blue]{hyperref}
\usepackage{amsthm}
\usepackage{amsmath, mathtools}
\usepackage[dvipsnames]{xcolor}
\usepackage{dsfont}
\usepackage[linesnumbered,ruled,vlined,noline]{algorithm2e}
\usepackage[noend]{algpseudocode}
\usepackage{xcolor}
\usepackage{soul}

\usepackage{changes}

\usepackage[capitalize]{cleveref}

\DeclareMathOperator{\sign}{sign}

\DeclareMathOperator{\gh}{gh}

\let\vec\mathbf 

\theoremstyle{definition}

\newcommand{\R}{\mathbb{R}}

\newcommand{\Z}{\mathbb{Z}}

\newcommand{\s}{\sigma}
\newcommand{\Lat}{\mathcal{L}}

\newcommand{\afunc}[1]{\operatorname{\mathsf{#1}}}

\begin{document}

\preprint{APS/123-QED}

\title{Beyond Ground States: Physics-Inspired Optimization of Excited States of Classical Hamiltonians}

\author{Erik Altelarrea-Ferré}
\affiliation{ICFO-Institut de Ciencies Fotoniques, The Barcelona Institute of Science and Technology, Castelldefels (Barcelona) 08860, Spain}

\author{Júlia Barberà-Rodríguez}
\affiliation{ICFO-Institut de Ciencies Fotoniques, The Barcelona Institute of Science and Technology, Castelldefels (Barcelona) 08860, Spain}

\author{David Jansen}
\affiliation{ICFO-Institut de Ciencies Fotoniques, The Barcelona Institute of Science and Technology, Castelldefels (Barcelona) 08860, Spain}

\author{Antonio Acín}
\affiliation{ICFO-Institut de Ciencies Fotoniques, The Barcelona Institute of Science and Technology, Castelldefels (Barcelona) 08860, Spain}
\affiliation{ICREA-Institucio Catalana de Recerca i Estudis Avançats, Lluis Companys 23, 08010 Barcelona, Spain}

\date{\today}
\begin{abstract}
We introduce excited local quantum annealing (ExcLQA), a classical, physics-inspired algorithm that extends local quantum annealing (LQA) to identify excited states of classical Ising Hamiltonians. LQA simulates quantum annealing while constraining the quantum state to remain in a product state and uses a gradient-based approach to find approximate solutions to large-scale quadratic unconstrained binary optimization problems. ExcLQA extends this framework by adding a penalty term in the cost function to target excited states, with a single hyperparameter that can be tuned via binary search to set the desired penalization level. We benchmark ExcLQA on fully connected Ising models with random interactions and on the shortest vector problem (SVP). The latter is a fundamental lattice problem underlying the security of many post-quantum cryptographic schemes, and its solution can be mapped to the first excited state of an Ising Hamiltonian. For the fully connected Ising models, we show that, on the tested instances, ExcLQA outperforms both a matrix-product-state-based method and simulated annealing. Notably, even when only a lower bound on the ground-state energy is provided, rather than the exact ground-state information required by these competing methods, ExcLQA still achieves superior performance. For the SVP, ExcLQA finds exact solutions for instances up to rank 46, and outperforms the Metropolis-Hastings algorithm in terms of solved ratio, number of shots, and approximation factor on the tested instances.
\end{abstract}

\maketitle

\section{Introduction \label{sec:intro}}

Many important scientific and technological challenges can be formulated as optimization problems, such as predicting protein structures~\cite{Anfinsen1973}, designing crystal structures with optimal properties~\cite{Oganov2006}, and scheduling tasks in production and computing environments~\cite{baker2018principles}. A subset of these admit representations in terms of a Hamiltonian energy landscape describing a system of interacting particles~\cite{glover_18}. In this formulation, the optimal solution corresponds to the ground state of the Hamiltonian, which can be targeted using computational methods that include both classical~\cite{Kirkpatrick1983,bowles,veszeli_21,schuetz_22,fioroni_25} and quantum~\cite{Farhi2001,Farhi2014,Peruzzo2014} techniques. Among these approaches, the \textit{physics-inspired} algorithms have gained particular prominence in recent years. This term refers to fully classical algorithms that build upon ideas from statistical physics and quantum many-body theory. Although the vast majority of these methods are heuristic and lack theoretical performance guarantees, several have achieved promising results in challenging optimization problems~\cite{mugel_22,tesoro_24,bowles,jansen_24,fioroni_25}, and they often provide \textit{high-quality} approximations to the ground state. 

An important class of optimization problems, however, falls outside the scope of ground-state search. In many cases, the global minimum of the cost function corresponds to a solution that is optimal in theory but impractical or undesirable in practice. As a result, the configuration of interest is not encoded in the ground state of the associated Hamiltonian but instead in one of its excited states. This situation can arise in constrained combinatorial optimization, including integer programming with exclusion rules~\cite{Wolsey1998}, and in scenarios where the lowest-cost solution fails to meet practical requirements. For instance, in facility location, the global minimum may place all facilities at the same site, which is often suboptimal~\cite{Daskin2013}, while in scheduling, minimizing cost alone can lead to inequitable workload allocation~\cite{Pinedo2016}. These challenges motivate the development of methods that enable more direct control over the energy level targeted during optimization and the systematic identification of alternative low-energy configurations, rather than relying solely on random fluctuations to escape the ground state. The central focus of this work is to develop methods that can target excited states in these challenging optimization problems. To motivate our contribution, we first review previously proposed strategies.

Classical strategies for finding low-energy excited states in Ising and related Hamiltonians can be broadly classified into three categories: stochastic samplers, deterministic enumerators, and landscape-reshaping schemes. Monte Carlo methods, including simulated annealing~\cite{Kirkpatrick1983}, parallel tempering~\cite{Hukushima1996}, and Metropolis–Hastings sampling~\cite{metropolis1953equation,robert2004metropolis}, are widely used to identify ground states and sample configurations near thermal equilibrium. Although they can occasionally produce excited configurations, they typically lack explicit mechanisms to reliably steer the optimization toward a chosen excitation level. Deterministic procedures, such as exact enumeration and branch-and-bound algorithms~\cite{Papadimitriou1982,Baxter1982}, are capable of systematically listing low-energy states for small systems but become intractable as dimensionality increases. Additional approaches have been developed to reshape the energy landscape, for example, by introducing penalty terms or constrained sampling protocols that discourage revisiting previously identified minima~\cite{Glover1998}. Other formulations instead introduce auxiliary variables to embed the target excitation into the ground-state manifold of an enlarged system~\cite{lucas2014ising}. Unlike purely combinatorial optimization techniques, the focus in these settings lies in accessing selected regions of the spectrum rather than exclusively identifying the ground state. However, the level of control over which excited state is ultimately reached remains limited.

In this work, we introduce excited local quantum annealing (ExcLQA), a classical physics-inspired algorithm for identifying low-energy excited states in classical Hamiltonians. ExcLQA builds on local quantum annealing (LQA), a technique introduced in Ref.~\cite{bowles} that simulates an adiabatic evolution toward an Ising Hamiltonian while constraining the state to remain a product state. This approach has been reported to yield \textit{high-quality} solutions to quadratic unconstrained binary optimization problems and was later extended to integer variables in Ref.~\cite{jansen_24} using an approach inspired by qudits. Here, we develop another extension to integer variables based on qubits. In addition, we incorporate a penalty term that increases the cost associated with low-energy configurations, raising not only the ground state but also excited states, with stronger penalization applied to states of lower energy. This penalty is controlled by a single hyperparameter, which allows the minimum of the cost function to be efficiently positioned near the target excitation level through binary search. While the specific penalty differs from the approach in Ref.~\cite{julia}, our method is motivated by the broader idea of penalizing low-energy configurations. We first benchmark ExcLQA on 20-qubit Ising models with random all-to-all interactions. When the exact ground-state energy is provided, ExcLQA attains a solved ratio of $0.87$ for the first excited state, significantly outperforming state-of-the-art QUBO-based and tensor-network-based approaches. When only a lower bound on the ground-state energy is available, these approaches are not applicable, as they require knowledge of the exact ground-state configuration, whereas ExcLQA still achieves a solved ratio of $0.70$. By tuning the penalty hyperparameter, our algorithm can also target several higher excited states with similarly high success rates. We then turn to the shortest vector problem (SVP), which consists of finding the shortest nonzero vector in a lattice. This problem is commonly formulated as finding the first excited state of an Ising Hamiltonian, with the ground state encoding the trivial zero vector~\cite{bose_hubbard,joseph_map}. The SVP is proven to be $\afunc{NP}$-hard under randomized reductions~\cite{Peikert_red}, which underscores its computational intractability in high dimensions and its central role in the security of lattice-based post-quantum cryptography (PQC)~\cite{Peikert_red,lwe}. This makes the SVP an appropriate benchmark for assessing the strengths and limitations of ExcLQA. In our benchmarks, ExcLQA consistently achieves a solved ratio above $0.67$ for ranks up to 39, and is able to find exact solutions on some instances up to rank 46. Despite these encouraging results, scaling beyond this range remains challenging, as hyperparameter fine tuning becomes increasingly demanding, and due to the heuristic nature of the method, there is no guarantee of correctness even with optimized parameters. Therefore, our method does not pose any threat to PQC schemes, as the cryptographically relevant lattices have ranks around 400.

The structure of the article is as follows. In Sec.~\ref{sec:prelim}, we establish notation and introduce the necessary theoretical concepts related to local quantum annealing, lattices, and the Hamiltonian formulation for the SVP. In Sec.~\ref{sec:methods}, we introduce our algorithm for excited states. In Sec.~\ref{sec:results}, we benchmark our algorithm on random Ising models and the SVP, offering a clear view of its performance and limitations. Finally, in Sec.~\ref{sec:conc}, we summarize our findings and outline potential directions for future research.

\section{Preliminaries} \label{sec:prelim}

Vectors are denoted in bold, such as $\vec{x}$, and the two-norm of a vector $\vec{x} \in \R^n$ is represented as $\norm{\vec{x}}$. Matrices are indicated by uppercase bold letters, e.g., $\vec{A}$. The $i$-th component of a vector $\vec{x}$ is written as $x_i$, and similarly, $A_{ij}$ denotes the $(i,j)$-th entry of a matrix $\vec{A}$. Operators are indicated with a hat, e.g., $\hat{H}$. 

\subsection{Local quantum annealing}\label{sec:excited-vqe}
In this subsection, we review local quantum annealing (LQA), introduced in Ref.~\cite{bowles}. LQA is a physics-inspired algorithm that efficiently finds high-quality solutions to quadratic unconstrained binary optimization (QUBO) problems on classical hardware. 

A QUBO problem is defined as
\begin{equation}\label{eq:qubo}
    \min_{\vec{x}\in\{0,1\}^n} \vec{x}^\top \mathbf{Q} \vec{x} + \vec{x}^\top \vec{a},
\end{equation}
where $\mathbf{Q}$ is an $n \times n$ real symmetric matrix and $\vec{a} \in \R^n$. Note that the linear term of \cref{eq:qubo} can be incorporated into the quadratic one by adding an extra variable with a fixed value of $1$. Thus, from hereon, we will assume $\vec{a}=\vec{0}$ without loss of generality. By mapping the binary variables $x_i\in \{ 0,1 \}$ to spin variables using the transformation $s_i=2x_i-1$, the  QUBO problem can be equivalently formulated as a quadratic unconstrained spin optimization (QUSO) problem. The solution of the QUSO problem can be equivalently obtained by finding the ground state of an Ising Hamiltonian:
\begin{equation}\label{eq:ising-ham}
    \hat{H}_z = \sum_{i,j=1}^n J_{ij}\hat \sigma_z^{(i)}\hat \sigma_z^{(j)},
\end{equation}
where $\hat \sigma_z^{(i)}$ is the Pauli $z$ operator on site $i$. This ground state can be accessed via quantum annealing, where the system is initialized in the ground state of a simple Hamiltonian and evolved slowly towards $\hat{H}_z$. If the evolution is performed slowly enough, the adiabatic theorem guarantees that the system remains in its ground state, ultimately yielding the global solution.

A typical formulation of quantum annealing consists of choosing $\hat{H}_x=\sum_{i=1}^n \hat \sigma_x^{(i)}$, where $\hat \sigma_x^{(i)}$ is the Pauli $x$ operator on site $i$, as initial Hamiltonian, so that the time-dependent Hamiltonian is given by
\begin{equation}
    \hat{H}(t) = t\kappa \hat{H}_z - (1-t)\hat{H}_x,\quad t\in [0,1],
\end{equation}
with $\kappa\in\R_{>0}$ controlling the relative strength of $\hat{H}_z$. The evolution begins from the state $|+\rangle^{\otimes n}$, the ground state of $\hat{H}(0)=-\hat{H}_x$. The target Hamiltonian, $\hat{H}(1)=\hat{H}_z$, is diagonal in the $Z$-basis, thus, at least one product state is part of its ground-state subspace. Leveraging the property that both the ground state of the initial and final Hamiltonian in $\hat{H}(t)$ can have a product form, LQA simulates quantum annealing by confining the evolution to product states, thus avoiding the exponential memory cost needed because of the entanglement buildup. This restriction suppresses entanglement and thus limits the exploration of the Hilbert space, breaking the conditions of the adiabatic theorem. As a consequence, the approach is heuristic and might potentially lead to suboptimal solutions. Yet, it has been observed that it offers good solutions, comparable or improved over the state of the art, for several QUBO problems~\cite{bowles}.

The states during the evolution are represented by the product ansatz

\begin{equation} \ket{\boldsymbol{\theta}} = \ket{\theta_1}\otimes\ket{\theta_2}\otimes\cdots\otimes\ket{\theta_n}, \end{equation}
with each single-qubit state defined as
\begin{equation}\label{eq:theta2} \ket{\theta_i} = \cos\frac{\theta_i}{2}\ket{+} + \sin\frac{\theta_i}{2}\ket{-}. \end{equation}
In this parametrization, the expectation values of the Pauli $z$ operators are
\begin{align}\label{eq:expval_theta2} \expval{\hat{\s}_z}{\theta_i} &=\sin\theta_i, & \expval{\hat{\s}_x}{\theta_i} &=\cos\theta_i. \end{align}
Thus, obtaining the ground-state energy of the Hamiltonian in Eq.~\eqref{eq:ising-ham} reduces to minimizing the expected energy over these product states
\begin{equation}
\expval{\smash{\hat{H}_z}}=\sum_{i,j=1}^{n}J_{ij}\sin\theta_i\sin\theta_j \, .
\end{equation}

The annealing process is simulated by discretizing the interval $[0,1]$ into $N$ points and updating the parameters via momentum-assisted gradient descent. The total cost function at time $t$ reads
\begin{align}\label{eq:total-cost-lqa}
    \mathcal{C}(t,\boldsymbol{\theta})
    =&t\kappa \sum_{i,j=1}^{n}J_{ij}\sin\theta_i\sin\theta_j \nonumber \\
    &- (1-t)\sum_{i=1}^{n}\cos\theta_i.
\end{align} 
To facilitate optimization, the angles $\theta_i$ are reparametrized as $\theta_i=\frac{\pi}{2}\tanh (w_i)$, where $w_i\in\R$ so that in the limits $w_i\to\pm\infty$, the states approach classical spin states $|0\rangle, |1\rangle$. This allows spin configurations to be extracted via $\sign(\vec{w})$.

To enhance efficiency, matrix-vector multiplications arising from the gradient of \cref{eq:total-cost-lqa} are delegated to a graphics processing unit. The initial weights $w_i$ are uniformly sampled from $[-f,f]\subset\R$, with $f$ being a tunable hyperparameter, introducing a stochastic element to the algorithm. Overall, LQA provides an efficient and scalable way to approximate solutions to QUBO problems by leveraging a restricted ansatz that captures the essential features of the ground state of an Ising Hamiltonian.

\subsection{Lattices}
A \textit{lattice} $\Lat$ is a discrete additive subgroup of $\R^d$. Equivalently, it is the set of all integer combinations of $n$ linearly independent vectors $\vec{b}_1, \dots, \vec{b}_n \in \R^d$, denoted as $\Lat(\vec{b}_1, \dots, \vec{b}_n) \coloneqq \left\{ \sum_{i=1}^n x_i \vec{b}_i \ \middle|\ x_i \in \Z\ \forall i \right\}$. We refer to $\vec{b}_1, \dots, \vec{b}_n$ as a \textit{basis} of $\Lat$, and to $n$ as the \textit{rank} of the lattice. When $n = d$, the lattice is said to be \textit{full-rank}. Bases are represented by matrices $\vec{B}$, with rows corresponding to the basis vectors, and we write $\Lat(\vec{B})$ to denote the lattice generated by $\vec{B}$. A lattice can have infinitely many bases, all related by unimodular transformations. The quality of a basis is often judged by how short and orthogonal its vectors are. The celebrated LLL, for Lenstra-Lenstra-Lovász, algorithm~\cite{LLL} was the first polynomial-time algorithm to compute an equivalent basis of guaranteed quality from any given basis.
For a given basis $\vec{B}$, the \textit{Gram matrix} is defined as $\vec{G} \coloneqq \vec{B}\vec{B}^\top$. The \textit{determinant} of the lattice is $\det(\Lat) \coloneqq \sqrt{\det(\vec{B}\vec{B}^\top)}$, where $\vec{B}$ is any basis of the lattice. The \textit{dual lattice} $\Lat^*$ is the set of vectors $\vec{x} \in \R^n$ such that $\vec{x} \cdot \vec{y} \in \Z$ for all $\vec{y} \in \Lat$. If $\vec{B}$ is a basis of $\Lat$, then $\vec{D} \coloneqq (\vec{B}\vec{B}^\top)^{-1} \vec{B}$ is a basis of $\Lat^*$.

A central problem in lattice-based cryptography is finding a shortest non-zero vector in a given lattice, known as the shortest vector problem (SVP)~\cite{lwe}. The length of such a vector is denoted by $\lambda_1(\Lat)$. By Minkowski's theorem~\cite{minkowskith}, this quantity is upper-bounded as $\lambda_1(\Lat) \leqslant \sqrt{n} \cdot \det(\Lat)^{1/n}$ for any full-rank lattice. This bound can be further refined using the \textit{Gaussian heuristic}, an asymptotic estimator for $\lambda_1(\Lat)$ in random lattices, defined as $\gh(\Lat) \coloneqq \sqrt{\frac{n}{2\pi e}} \cdot \det(\Lat)^{1/n}$.
The best classical algorithms for the SVP are based on two main approaches: enumeration and sieving. These algorithms run in $n^{\mathcal{O}(n)}$, $2^{\mathcal{O}(n)}$ time, and require polynomial and exponential memory, respectively, where $n$ is the rank of the lattice~\cite{complexity_svp}.

A relaxed variant of the SVP is the \textit{approximate shortest vector problem} ($\gamma$-SVP), where, given an approximation factor $\gamma \in \mathbb{R}_{\geq 1}$, the goal is to find a vector $\vec{v} \in \Lat \setminus \{\vec{0}\}$ such that $\norm{\vec{v}} \leqslant \gamma \cdot \lambda_1(\Lat)$. Notably, in the $p$-norm, the $\gamma$-SVP is proven to be $\afunc{NP}$-hard under randomized reductions for any $\gamma < 2^{1/p}$~\cite{Micciancio01svp}.

\subsection{Hamiltonian formulation for the SVP}\label{sec:ham_formulation}
Given a lattice of rank $n$, $\Lat(\vec{B})$, where the rows of $\vec{B}$ are the vectors $\vec{b}_i \in \R^d$, any lattice point $\vec{v} \in \Lat(\vec{B})$ can be written as 
\begin{equation}
\vec{v}=\vec{xB}=x_1\vec{b}_1+\dots+x_n\vec{b}_n, \quad x_i\in \Z\ \forall i.
\end{equation}
To optimize the formulation, it is standard to work with the norm squared and seek the non-zero vector that minimizes it.

Let $\vec{G}\coloneqq\vec{B}\vec{B}^\top$ be the Gram matrix of the basis vectors $\vec{b}_1, \dots, \vec{b}_n$. The squared norm of a lattice vector $\vec{v}=\vec{x}\vec{B}$ can be expressed in terms of its coefficients $\{x_i\}_{i=1}^n$ and the entries of the Gram matrix as \begin{equation}\label{eq:norm} 
\norm{\vec{v}}^2 = \sum_{i,j=1}^n x_i x_j G_{ij}. 
\end{equation} 
Hence, solving the SVP reduces to finding a vector $\vec{x}\in~\Z^n\setminus \{\vec{0}\}$ that minimizes \cref{eq:norm}. 

To obtain a Hamiltonian representation of this problem, we follow the approach of Ref.~\cite{not_so_adiabatic}, mapping \cref{eq:norm} to the following Hamiltonian: \begin{equation}\label{eq:ham_svp}
    \hat{H}_z=\sum_{i,j=1}^n\hat{Q}^{(i)}\hat{Q}^{(j)}G_{ij}  .
\end{equation} The Hamiltonian in \cref{eq:ham_svp} is designed such that the operators $\hat{Q}^{(i)}$, when applied to a string of qubits, yield the integer values associated to the vector coefficients $\{x_i\}_{i=1}^n$. Hence, the action of the Hamiltonian over an eigenstate is
\begin{equation}
\hat{H}_z\ket{\psi_{\vec{v}}}=\norm{\vec{v}}^2\cdot\ket{\psi_{\vec{v}}}, 
\end{equation}
where $\ket{\psi_{\vec{v}}}$ encodes the lattice vector $\vec{v}\in~\Lat(\vec{B})$, and its corresponding eigenvalue is the squared length of this vector, $\norm{\vec{v}}^2$. The ground-state energy of Hamiltonian \cref{eq:ham_svp} is equal to zero, with the ground state corresponding to the trivial solution defined by the zero vector. In contrast, the shortest nontrivial vector corresponds to the first excited state. The operators $\hat{Q}^{(i)}$ in \cref{eq:ham_svp} were introduced in Ref.~\cite{joseph_map}. For a detailed explanation of their expressions and how they act on the eigenstates, refer to App.~\ref{app:qudits-operators}.

\section{Method \label{sec:methods}}
Many heuristic methods are primarily developed to identify ground-state configurations and can occasionally converge to excited states in practice. However, they often lack explicit mechanisms to reliably steer the optimization toward a chosen excitation level~\cite{metropolis1953equation,robert2004metropolis,Kirkpatrick1983,Hukushima1996}. In this section, we introduce excited local quantum annealing (ExcLQA), an extension of local quantum annealing that provides enhanced control over this process in Ising Hamiltonians. ExcLQA reshapes the optimization landscape by adding an inverse-energy penalty term that increases the cost as the energy decreases, thereby raising not only the ground state but also excited states, with stronger penalization applied to lower-energy configurations. By tuning the penalization hyperparameter, the minimum of the cost function can be shifted close to the desired excitation level through an efficient binary-search procedure.

To formulate this strategy, and motivated by the general concept of energy penalization introduced in Ref.~\cite{julia}, we define the penalized cost function as
\begin{equation}\label{eq:fin_cost}
    E_\text{F}(\boldsymbol{\theta})\coloneqq \expval{\smash{\hat{H}_z}}+\frac{\alpha}{\expval{\smash{\hat{H}_z}}}, 
\end{equation}
where $\alpha\in\R_{>0}$ is a tunable penalization hyperparameter. Note that the cost function in~\cref{eq:fin_cost} does not correspond to the energy of any physical Hamiltonian. To guarantee that the penalization increases as the energy decreases, the ground-state energy of $\hat{H}_z$ must be nonnegative. If this condition is not satisfied, ExcLQA can still be applied by shifting the spectrum upward through the addition of an appropriate constant offset to $\hat{H}_z$. In practice, this constant can be determined by computing a lower bound to the ground-state energy, for instance, using semidefinite programming relaxations~\cite{goemans1995improved,lasserre_2001} or the Anderson bound~\cite{Anderson1951}. The resulting Hamiltonian is guaranteed to have a nonnegative energy spectrum.

Following the structure of adiabatic quantum evolution, we define $E_{\text{I}}(\boldsymbol{\theta})\coloneqq-\expval{\smash{\hat{H}_x}}$ and combine these functions into the total cost as
\begin{equation}\label{eq:total_cost}
    E_\text{Total}(t,\boldsymbol{\theta})\coloneqq (1-t)E_\text{I}(\boldsymbol{\theta})+t^{\beta}\kappa E_\text{F}(\boldsymbol{\theta}).
\end{equation}
Here, $t \in [0, 1]$ is the evolution parameter, $\kappa \in \R_{>0}$ controls the strength of the final cost function, and $\beta \in \R_{>0}$ allows heuristic tuning of the interpolation schedule. In our numerical experiments presented in Sec.~\ref{sec:bench}, we set $\beta=3.8$, selected empirically to enhance performance. We simulate the adiabatic evolution of~\cref{eq:total_cost} by discretizing $[0,1]\subset~\R$ into $N$ points and updating the parameters at each point using the stochastic gradient descent (SGD)~\cite{SGD} implementation in PyTorch~\cite{pytorch}. Here, $N$ controls the resolution of the interpolation, and should not be interpreted as a physical annealing time. Once chosen large enough to ensure stable convergence, we observed that further increases have a limited impact on the solution quality, while significantly increasing computational cost.

Algorithm~\ref{alg_gd} summarizes our approach. Here, the function $g$ represents a generic parameter update of~\cref{eq:total_cost} using SGD, which may incorporate additional inputs such as learning rate and momentum. 
\vspace{2mm}
\begin{algorithm}[H]\label{alg_gd}
\KwIn{$H_z$: Ising Hamiltonian; $N$: total points; $\alpha$: penalization hyperparameter.}
\KwOut{$\vec{s}$: excited state.}
$\boldsymbol{\theta}_0\gets \text{Initial parameters}$\\
 \For{$i=1,\dots, N$}
 {$\boldsymbol{\theta}_i\leftarrow  g(\boldsymbol{\theta}_{i-1},\nabla_{\boldsymbol{\theta}} E_{\text{Total}}(i/N, \boldsymbol{\theta}_{i-1}))$}
 $\vec{s}\gets\textsc{Decode}(\boldsymbol{\theta}_N)$\\
 \Return{$\vec{s}$}
 \caption{ExcLQA}
\end{algorithm}
\vspace{2mm}

\begin{figure}[ht]
    \includegraphics[width=\columnwidth]{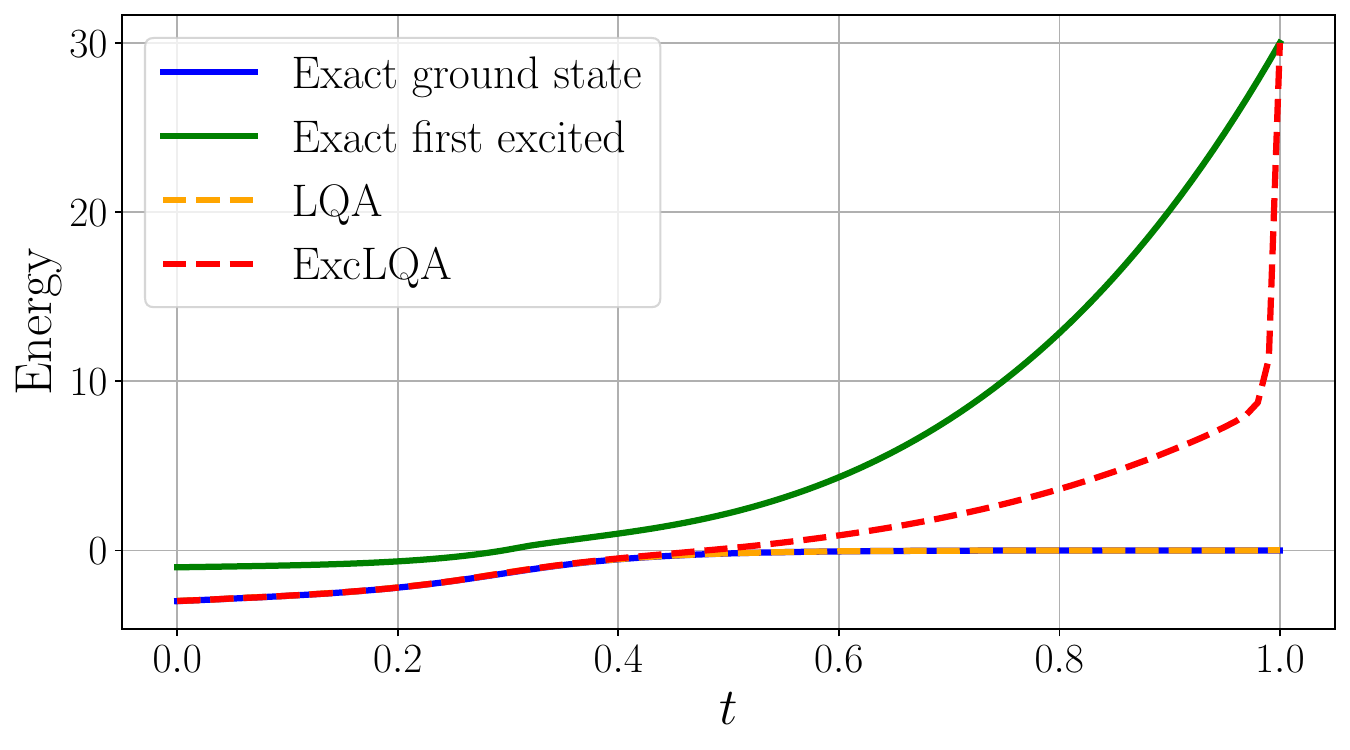}
    \caption{\justifying Comparison of the energy minimization for LQA and ExcLQA as a function of the evolution parameter $t$. The orange dashed line represents the energy evolution of LQA, which outputs the ground state. In contrast, the red dashed line shows the energy evolution of ExcLQA, where the hyperparameter $\alpha$ in \cref{eq:fin_cost} is tuned to target a first excited state.}
    \label{fig:penalising}
\end{figure}

\begin{figure}[ht]
    \centering
    \includegraphics[width=\columnwidth]{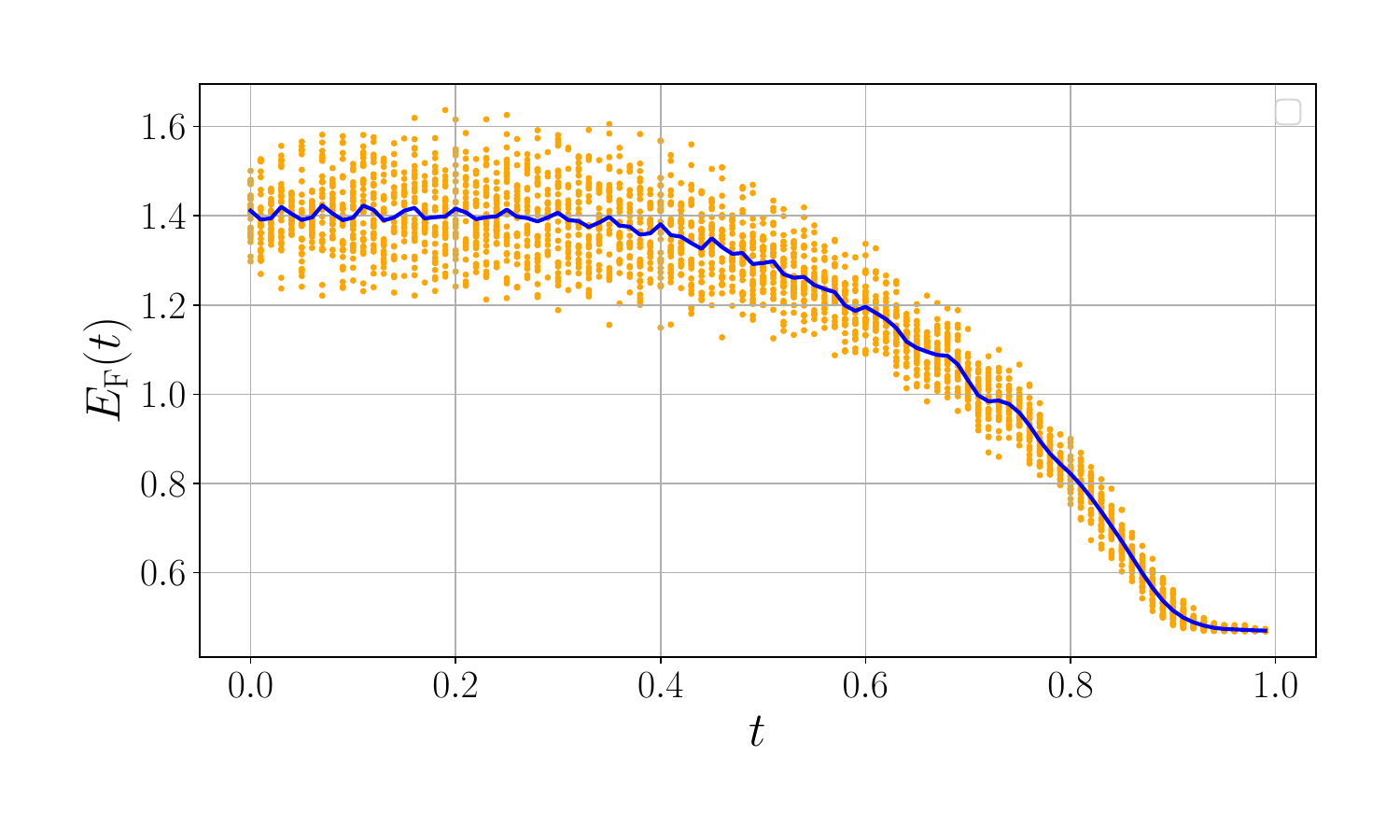}
    \caption{\justifying Evolution of the final cost function as a function of the interpolation parameter $t$. The orange points indicate the values of the cost function at each step, while the solid blue line represents the average over 40 shots.}
    \label{fig:energy}
\end{figure}

To illustrate the dynamics of our algorithm, in \cref{fig:penalising}, we plot the value of \cref{eq:total_cost} at each interpolation step $t$, where the penalization hyperparameter has been tuned to facilitate convergence towards the first excited state. The hyperparameters used in Figs.~\ref{fig:penalising} and~\ref{fig:energy} are detailed in App.~\ref{app:hyperparam}. The dashed yellow and red lines represent, respectively, the time evolution of the ground‑state and first‑excited‑state energies of the following Ising Hamiltonian: \begin{align}
\hat{H}_z=&93\hat{\mathds{1}}+18\hat{\s}_z^{(1)}+30\hat{\s}_z^{(2)}+24\hat{\s}_z^{(3)}\notag \\ 
&+3\hat{\s}_z^{(1)}\hat{\s}_z^{(2)}-24\hat{\s}_z^{(2)}\hat{\s}_z^{(3)}.
\end{align}
This Hamiltonian was obtained from applying the mapping of Sec.~\ref{sec:ham_formulation} to a random SVP instance of rank $n=3$. The simulation results are compared to the exact energy values obtained via diagonalization, shown as solid lines for the ground state (blue) and the first excited state (green). Figure~\ref{fig:penalising} shows that LQA closely tracks the exact ground state energy curve, while ExcLQA transitions to the energy of the first excited state and converges to the first excited state.

To illustrate how ExcLQA minimizes the final cost function in practice, Fig.~\ref{fig:energy} shows the evolution of $E_\text{F}$ as a function of the interpolation parameter $t$. The simulations are performed for a 35‑spin Ising Hamiltonian with zero ground‑state energy. The results empirically demonstrate the minimization of the final cost function, mimicking a quantum adiabatic evolution, and obtaining a first excited state.

It is important to remark that our algorithm avoids introducing auxiliary variables to recast the target excited state into an effective ground state, as is commonly done in penalty-based formulations~\cite{lucas2014ising}, and does not require any additional preprocessing costs. Moreover, the hyperparameter $\alpha$ in \cref{eq:fin_cost} can be efficiently tuned via binary search to select the desired penalization level.

\section{Results} \label{sec:results}
In this section, we benchmark ExcLQA on two problems. In Sec.~\ref{sec:fcIsing}, we consider fully connected Ising Hamiltonians with random interactions. In Sec.~\ref{sec:svp}, we introduce the SVP setting considered in our experiments, where a solution can be mapped to a first excited state following the approach described in Sec.~\ref{sec:ham_formulation}, and in Sec.~\ref{sec:bench}, we present the corresponding benchmark results.

\subsection{Random fully connected Ising Hamiltonians}\label{sec:fcIsing}
We first test ExcLQA on 500 instances of fully connected 20-spin Ising Hamiltonians with random all-to-all couplings. For each instance, we sample a matrix $\tilde{\vec{J}}$ with entries drawn uniformly from $[-1,1]\subset\R$, and define the symmetric coupling matrix as $\vec{J}\coloneqq\tilde{\vec{J}}+\tilde{\vec{J}}^{\top}$, with zero diagonal. This coupling matrix defines an Ising Hamiltonian of the form given in Eq.~\eqref{eq:ising-ham}.

\begin{figure}[ht]
    \begin{subfigure}[a]{1\columnwidth}
        \includegraphics[width=\textwidth]{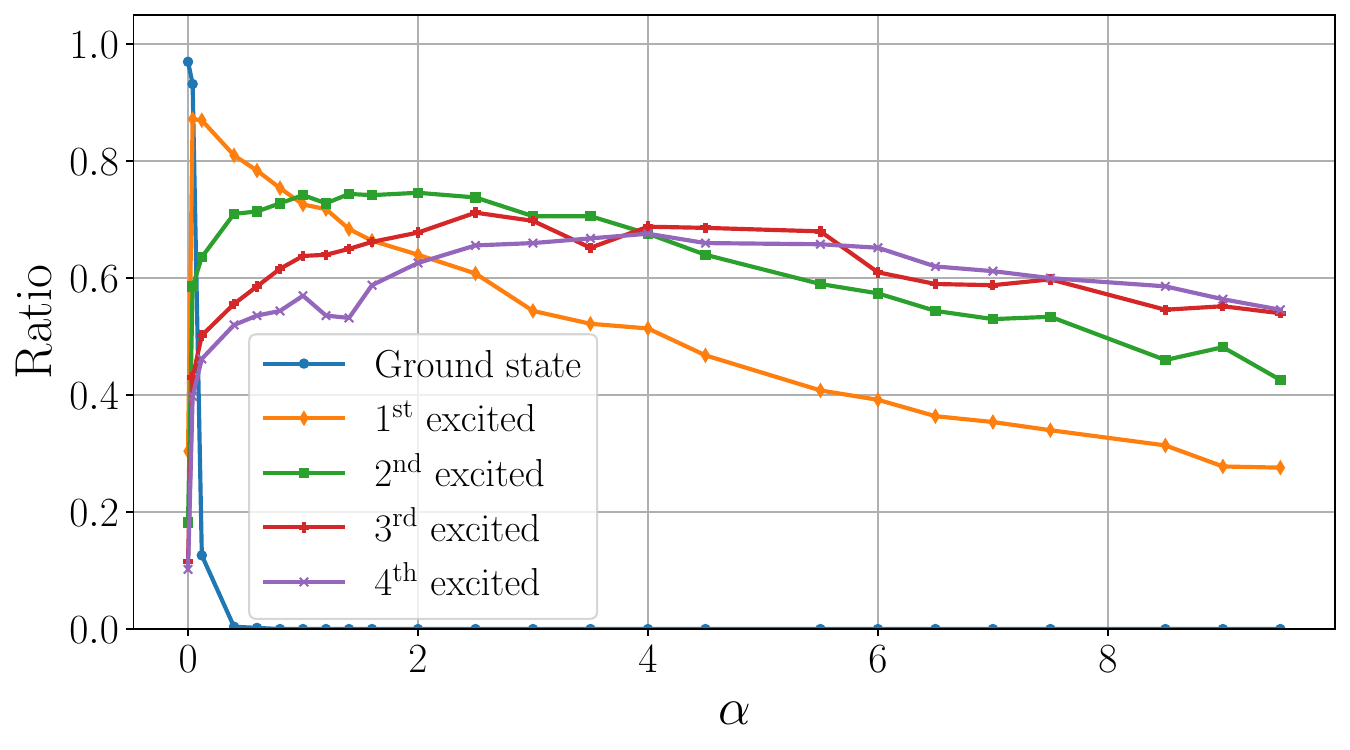}
        \caption{Exact ground-state energy shift.}
        \label{fig:ising_true_energy}
    \end{subfigure}
    \begin{subfigure}[b]{1\columnwidth}
        \includegraphics[width=\textwidth]{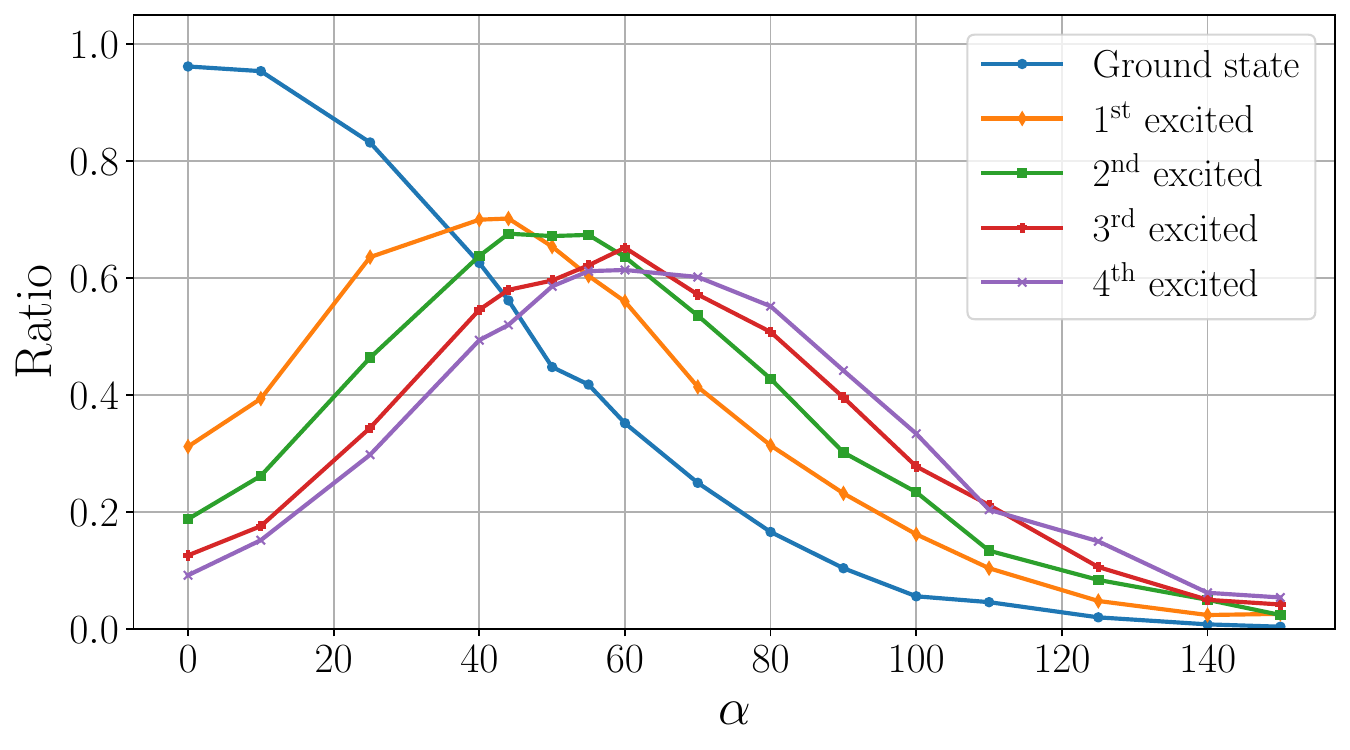}
        \caption{SDP-based lower-bound energy shift.}
        \label{fig:sdp}
    \end{subfigure}
    \caption{\justifying Ratio of problem instances for which ExcLQA visits the ground state and the first four excited energy levels as a function of the penalty parameter $\alpha$, evaluated over $500$ random Ising Hamiltonians. For each instance and value of $\alpha$, the algorithm is run with $50$ independent shots, and an energy level is considered visited if at least one shot converges to it.}
    \label{fig:3}
\end{figure}

Figure~\ref{fig:ising_true_energy} shows the ratio of instances for which ExcLQA converges to the ground state and to the first four excited states as a function of the penalty hyperparameter $\alpha$. In this experiment, the exact ground-state energy is provided to ExcLQA to shift the spectrum as required by the penalized objective. At $\alpha=0$, ExcLQA reduces to ground-state search and reaches the ground state with a ratio of $0.97$. As $\alpha$ increases, the algorithm is progressively steered toward higher excited states: the first excited state attains a maximum ratio of $0.87$ at $\alpha=0.04$, while the second, third, and fourth excited states attain their maximum ratios at $0.74$, $0.71$, and $0.67$ at $\alpha=2$, $2.5$, and $4$, respectively. Overall, these results show that, by properly tuning $\alpha$, ExcLQA can be utilized to target higher excited states, though this becomes increasingly more difficult higher in the spectrum.

To further benchmark ExcLQA, we compare it against a standard QUBO-based penalty strategy. Representing a spin configuration in its binary form, we denote by $\mathbf{x}\in\{0,1\}^n$ a candidate solution and by $\mathbf{x}_0$ the binary representation of the ground state, which is required as an explicit input to the method. To suppress the ground state and favor excited configurations, we add a quadratic penalty that enforces a positive Hamming distance from $\mathbf{x}_0$, as commonly done to penalize undesired solutions in Ising and QUBO formulations (see, e.g., Ref.~\cite{lucas2014ising}). The resulting cost function is \begin{equation}E_{\mathrm{QUBO}}(\mathbf{x}) \coloneqq E(\mathbf{x}) + \lambda\bigl(d(\mathbf{x},\mathbf{x}_0)-d_0\bigr)^2,\end{equation} where $d$ denotes the Hamming distance, $d_0\in[1,n]\cap\mathbb{Z}$ and $\lambda\in\mathbb{R}_{>0}$ are tunable hyperparameters controlling the target distance and the penalty strength, respectively. We solve the resulting QUBO instances using classical simulated annealing implementation of PyQUBO~\cite{pyqubo}, tuning $\lambda$ and $d_0$. The best configuration, given by $d_0=1$ and $\lambda=50$, attains a solved ratio of $0.55$ for the first excited state over the 500 instances, compared with $0.87$ achieved by ExcLQA under the same benchmark conditions.
   
Additionally, we benchmark ExcLQA against a tensor-network-based approach by employing a matrix-product state ansatz and running a density-matrix renormalization group (DMRG) calculation~\cite{PhysRevLett.69.2863} on the Ising Hamiltonian, while enforcing orthogonality to the ground state, which is a standard procedure for targeting excited states in quantum many-body physics~\cite{Stoudenmire_12}. In our implementation, we initialize DMRG from a random state with bond dimension $100$ and optimize the modified Hamiltonian \begin{equation} \hat H^{\prime}\coloneqq \hat H + \lambda^{\prime}\ket{\psi_0}\bra{\psi_0},\end{equation} where $\hat H$ is the Ising Hamiltonian, $\ket{\psi_0}$ denotes the ground state, and  $\lambda^{\prime}\in\mathbb{R}_{>0}$ is a tunable hyperparameter. In the best-performing configuration, with $\lambda^{\prime}=10$, the method converges to the first excited state with a ratio of $0.51$ across the 500 instances, compared with $0.87$ for ExcLQA, while converging to the true ground state ($\ket{\psi_0}$) with a ratio of $0.02$. All simulations were carried out using ITensor~\cite{itensor}.

Lastly, we demonstrate that ExcLQA does not require knowledge of the true ground state to remain effective, which is a crucial advantage since identifying the ground state itself can be an equally challenging task. In Fig.~\ref{fig:sdp}, we report the same analysis as in Fig.~\ref{fig:ising_true_energy}, but instead of shifting the spectrum using the exact ground-state energy, we shift it using a lower bound obtained from a semidefinite programming (SDP) relaxation~\cite{goemans1995improved,lasserre_2001}. In our implementation, we consider a first-order relaxation retaining only constant and linear monomials. We formulate the resulting SDP in CVXPY~\cite{diamond2016cvxpy,agrawal2018rewriting} and solve it with the SCS conic solver~\cite{odono2016scs}. Across the 500 instances, the relative difference between the true ground-state energy $E_{\mathrm{gs}}$ and the SDP lower bound $E_{\mathrm{lb}}$, 
\begin{equation}
\Delta_E=\frac{E_{\mathrm{gs}}-E_{\mathrm{lb}}}{\abs{E_{\mathrm{gs}}}},
\end{equation}
has a mean value of $9.68\times10^{-2}$ with a standard deviation of $3.80\times10^{-2}$. Despite this relatively loose yet inexpensive bound, ExcLQA continues to identify multiple excited states with high success rates, as shown in Fig.~\ref{fig:sdp}. In particular, the maximum solved ratios for the ground, first, second, third, and fourth excited states are $0.962$, $0.702$, $0.676$, $0.652$, and $0.614$, attained at $\alpha=0$, $44$, $44$, $60$, and $60$, respectively. These values of $\alpha$ are much larger than those in Fig.~\ref{fig:ising_true_energy} because using the exact ground-state energy shifts the ground state to zero, where the penalty term becomes dominant. In contrast, when the SDP lower bound is used, the spectral shift does not bring the ground state to zero, but leaves it at a positive shifted energy. Consequently, larger values of $\alpha$ are required to penalize it relative to the excited states. The hyperparameters used for the ExcLQA simulations were $N=200,\kappa=5,f=0.2,\mu=0.999,\eta=0.999$.

Beyond providing a benchmark across different excitation levels, Fig.~\ref{fig:3} also clarifies how the penalty hyperparameter $\alpha$ in \cref{eq:fin_cost} should be tuned in practice. For a fixed energy level, we observe a simple and robust empirical pattern: as $\alpha$ increases, the success ratio rises monotonically, reaches a well-defined maximum, and then decreases monotonically as the penalty becomes too strong and pushes the optimization toward higher-energy configurations. Importantly, the maximum is typically attained over a relatively broad range of $\alpha$ values rather than at a sharply peaked optimum, indicating that the method does not rely on delicate fine-tuning. This unimodal dependence is observed across all tested instances and excitation levels. As a result, the optimal value of $\alpha$ can be efficiently identified using a simple bracketing or binary-search strategy, without requiring fine-grained grid searches or additional hyperparameters.

\subsection{Shortest Vector Problem}\label{sec:svp}
We consider several SVP instances in $q$-ary lattices, which are particularly significant in cryptography because of their enhanced memory representation and their role in average-case to worst-case reductions. For example, finding short vectors in the dual of a random $q$-ary lattice is as hard as finding short vectors in any lattice~\cite{red_lat}.

The basis of a $q$-ary lattice is given by a matrix of the form 
\begin{equation}\label{eq:q_ary_B}
    \begin{pmatrix}
        \vec{I}_{d-k} & \widetilde{\vec{A}}\\
        \vec{0} & q\cdot\vec{I}_k
    \end{pmatrix}\in \Z^{d\times d},
\end{equation}
where $\vec{I}_k\in\Z^{k\times k}$ denotes the $k\times k$ identity matrix. Following the approach outlined in Ref.~\cite{Albrecht_2023}, we set $q =~2^{16} + 1$, $d = 180$, $k = 90$ and sample $\widetilde{\vec{A}}$ uniformly from $\Z_q^{(d-k)\times k}$ to generate a matrix in the form of \cref{eq:q_ary_B}. To evaluate the performance of ExcLQA, we first generate $d$-dimensional lattices. Since the LLL algorithm cannot directly solve the SVP in $d=180$ dimensional lattices, we use it as a preprocessing step to obtain a better-quality basis. Next, we extract an $n$-rank sublattice from the preprocessed basis by selecting the first $n$ rows. This $n\times d$ matrix serves as the input for our algorithm, which then aims to find a shortest vector in the generated sublattice.

It is important to note that while the described procedure effectively improves the quality of the basis used as input for our algorithm, it does not reduce the size of the search space. Consequently, solving the SVP in these bases remains a significant challenge, and by sampling random initial parameters, our experiments simulate real instances on a reduced scale.

Determining an appropriate search space is challenging: if the search space is overly large, the likelihood of converging to a solution decreases, whereas if it is overly restricted, it may exclude true solutions and result in a failure rate equal to one for any algorithm. This challenge arises directly when implementing the Hamiltonian in \cref{eq:ham_svp} within ExcLQA, where it is necessary to establish bounds on the spectrum of the $\hat{Q}^{(i)}$ operators, i.e., on each vector coefficient $x_i$. Assuming that $\norm{x_1\vec{b}_1+\dots+x_n\vec{b}_n}\leqslant A$, Ref.~\cite[Lemma 1]{Albrecht_2023} proves that $|x_i|\leqslant A\cdot\lVert\vec{d}_i\rVert$, where $\vec{d}_i$ are the rows of $\vec{D}$, a basis of the dual lattice $\Lat^*$. Under the Gaussian heuristic, setting $A=\gh(\Lat)$ would provide an upper bound for each $|x_i|$, but this leads to an excessively large search space that hampers convergence. 

\begin{figure}[ht]
    \centering
    \includegraphics[width=\columnwidth]{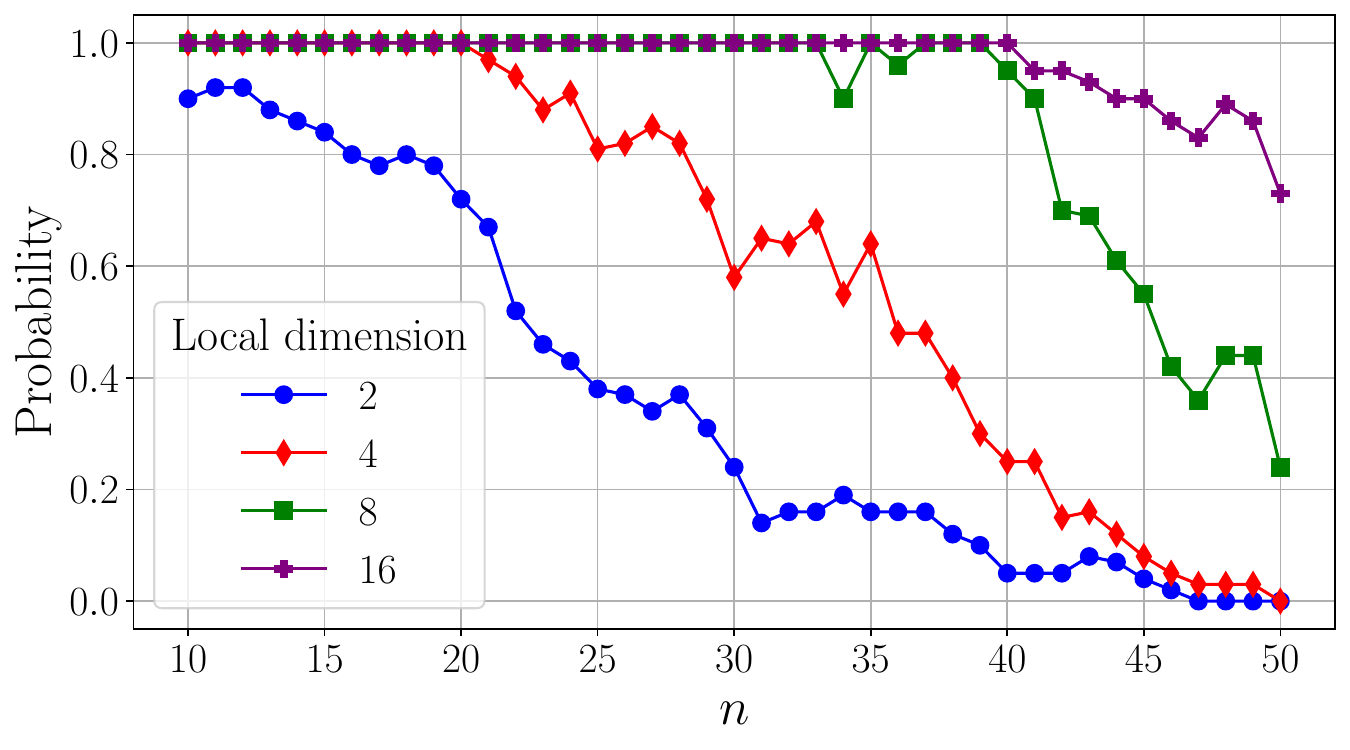}
    \caption{\justifying Probability of including a solution in the search space as a function of the rank, $n$, using naive qubit mappings. We generated $100$ instances of the problem for each $n$ and computed the probability of including a solution for the SVP in the search space defined by different local-dimensional qudits.}
    \label{fig:search_space_prob}
\end{figure}

This motivates the adoption of minimal local dimensions. By employing the binary-encoded qudits mapping from Ref.~\cite{joseph_map}, the search space when the local dimension is set to $2^k$ becomes
\begin{equation}\label{eq:search_space_k}
\{x_1\vec{b}_1+\dots+x_n\vec{b}_n\mid x_i\in [-2^{k-1},2^{k-1}-1]\cap\Z\ \forall i\}.
\end{equation}
In \cref{fig:search_space_prob}, we illustrate how the local dimension affects the probability that a shortest non-zero vector is contained within the search space. This probability is shown as a function of the sublattice rank for several local dimensions. Here, the local dimension refers to the number of bits used to represent each vector coefficient $x_i$. Figure~\ref{fig:search_space_prob} shows that for local dimensions of $2$ and $4$ the probability decreases approximately linearly in the lattice rank, whereas for $8$ and $16$ it remains over $0.9$ for ranks between $10$ to $40$ before starting to decrease linearly. In particular, for a local dimension of $2$, the probability of including a solution in the search space falls below $0.05$ for lattice ranks greater than $40$, limiting its utility to ranks up to $39$. 

\subsection{Benchmark}\label{sec:bench}
For our simulations, $100$ sublattices were generated for each sublattice rank $n$ using the procedure described in Sec.~\ref{sec:svp}, and a shortest non-zero vector for each sublattice was computed using the classical enumeration implementation from the library in Ref.~\cite{fplll}. Following Refs.~\cite{Albrecht_2023,prokop2025heuristictimecomplexitynisq}, we then restricted our analysis to those instances in which the solution was expressible within the search space defined in \cref{eq:search_space_k} with the local dimension set to 2, i.e., each vector coefficient is either $-1$ or $0$. We report the number of these instances for each lattice rank $n$ in Tab.~\ref{table:performance_exclqa}, together with the corresponding solved instances, average number of shots, and approximation factors for both variants of ExcLQA.

\begin{figure}[ht]
    \begin{subfigure}[a]{1\columnwidth}
        \includegraphics[width=\textwidth]{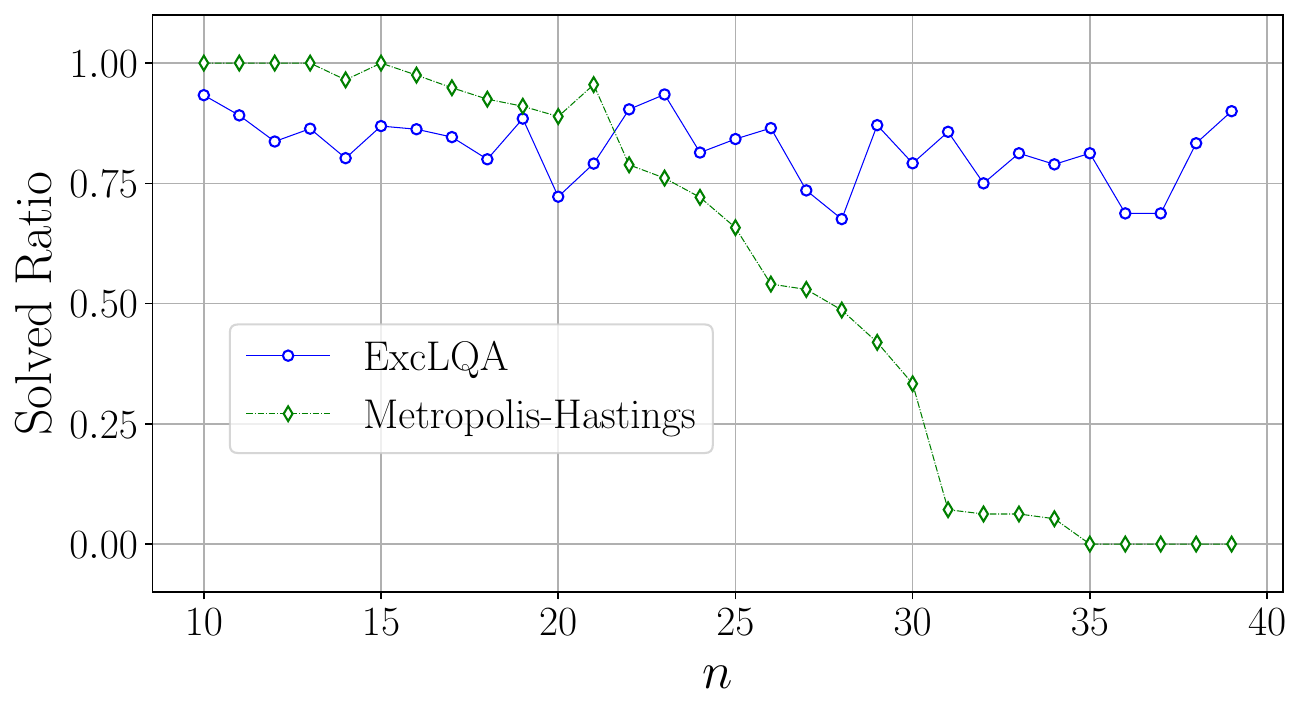}
        \caption{Success rate of instances solved using our method.}
        \label{fig:solved_ratio}
    \end{subfigure}
    \begin{subfigure}[b]{1\columnwidth}
        \includegraphics[width=\textwidth]{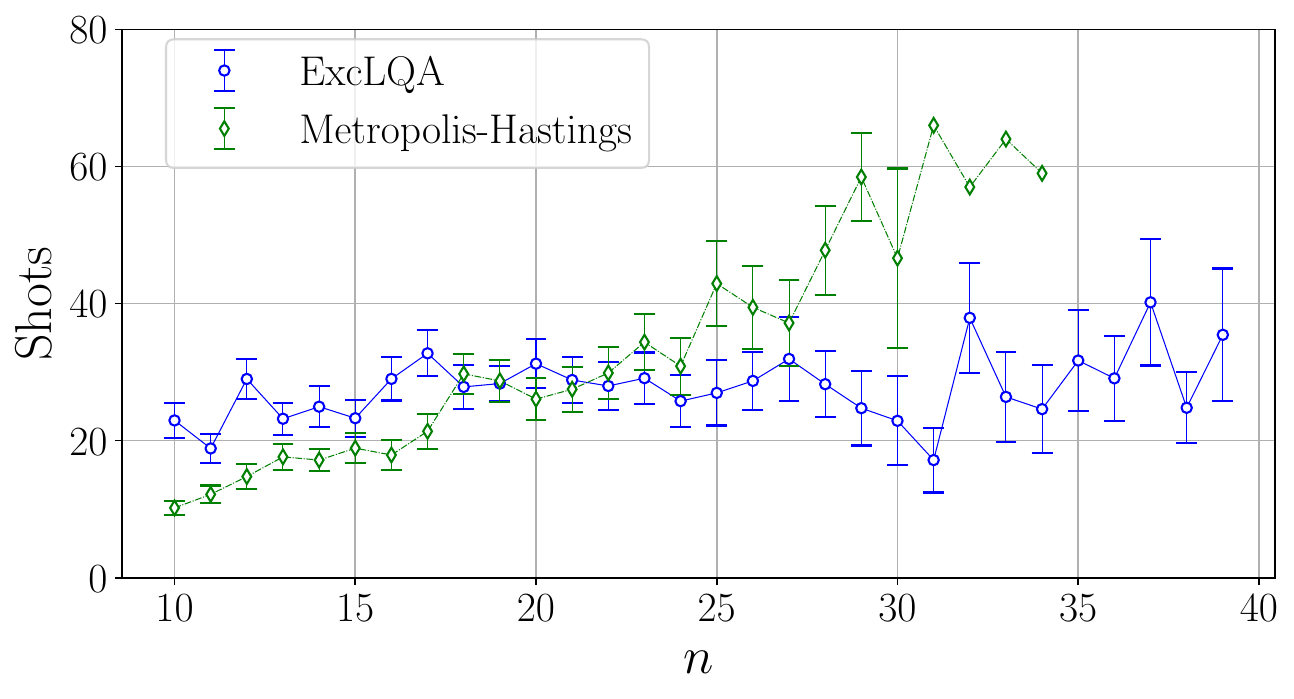}
        \caption{Averaged number of shots until success.}
        \label{fig:shots}
    \end{subfigure}
    \caption{\justifying Results of ExcLQA and Metropolis-Hastings for solving the SVP. We generated $100$ instances of the problem for each rank $n$ and restricted to those whose solutions were contained within the search space in \cref{eq:search_space_k} defined by setting the local dimension to 2. Each method was run using $N=100$ points/iterations.}
    \label{fig:results}
\end{figure}

For comparison, we benchmark against the Metropolis-Hastings algorithm~\cite{metropolis1953equation, metropolis}, a standard physics-inspired method that efficiently samples low-energy configurations, and which we describe in detail in App.~\ref{app:metropolis}. To target excited states using the Metropolis-Hastings algorithm, we introduce the same penalty term as in ExcLQA. In \cref{fig:solved_ratio}, we show the performance of ExcLQA relative to this baseline, which we apply to minimize the cost function in \cref{eq:fin_cost}. We plot the solved ratio, defined as the proportion of instances where a vector of length $\lambda_1(\Lat)$ (i.e., a shortest non-zero vector) was identified, as a function of $n$. The solved ratio for ExcLQA remains relatively stable with increasing sublattice rank, staying above $0.675$ and averaging approximately $0.822$. This empirically demonstrates effective performance despite the growing complexity. In contrast, the solved ratio for the Metropolis-Hastings algorithm exhibits a linear decay as the sublattice rank increases.

Additionally, \cref{fig:shots} shows the average number of shots required by both algorithms to find a solution, where a shot refers to a single execution of each algorithm with a random perturbation of the initial state. In all simulations, the maximum number of shots was limited to 100. ExcLQA consistently requires fewer than 40 shots on average, indicating robustness across different sublattice ranks, whereas the Metropolis-Hastings algorithm shows a linear increase in the number of shots as the sublattice rank grows.

\begin{figure}[ht]
    \centering
    \includegraphics[width=\columnwidth]{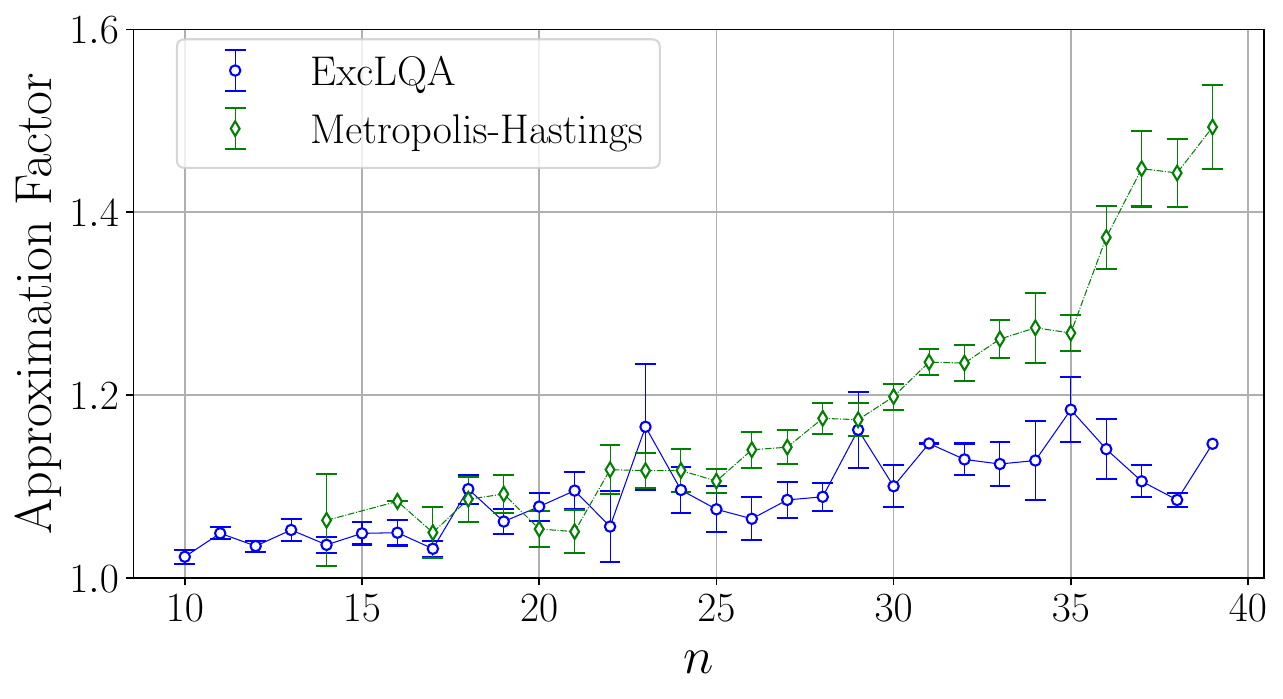}
    \caption{\justifying Average approximation factor computed among the instances where each method failed to solve the SVP within 100 shots. The approximation factor is defined as $\norm{\vec{v}} / \lambda_1(\Lat)$, where $\vec{v}$ is the sublattice vector identified by each method and $\lambda_1(\Lat)$ denotes the length of a shortest non-zero vector.}
    \label{fig:aprox_factor}
\end{figure}

In \cref{fig:aprox_factor}, we plot the average of the best approximation factor obtained from all shots for each sublattice rank $n$ in instances where the SVP was not solved within 100 shots. The approximation factor is defined as $\gamma = \norm{\vec{v}}/\lambda_1(\Lat)$, where $\vec{v}$ represents the sublattice vector found by each algorithm, and $\lambda_1(\Lat)$ denotes the length of a shortest non-zero vector. ExcLQA consistently finds good approximations for the $\gamma$-SVP, with $\gamma$ increasing only slightly with $n$, likely reflecting the increased challenge of finding short vectors in higher dimensions. Notably, since the approximation factor remains below $\sqrt{2}$, our method effectively addresses instances of this $\afunc{NP}$-hard problem in the two-norm~\cite{Micciancio01svp}. In contrast, the approximation factor for the Metropolis-Hastings algorithm increases more abruptly with sublattice rank, surpassing the $\sqrt{2}$ threshold at rank 37.

To test ExcLQA at larger sublattice ranks, we set the local dimension of the search space in \cref{eq:search_space_k} to 4, as Fig.~\ref{fig:search_space_prob} indicates that for a local dimension of 2, the probability of including a solution is below $0.05$ for ranks above 39. This allows ExcLQA to identify lattice vectors whose coefficients are $-2,-1,0$ or $1$, but also doubles the size of the search space, increasing the number of spins of the Ising Hamiltonian from $n$ to $2n$ and significantly complicating fine-tuning. Despite these challenges, ExcLQA was able to solve some instances even at higher ranks. The following fractions represent the number of solved instances over the total whose solution was contained in the search space: $n=43:\ 4/16,\ n=44:\ 3/12,\ n=~45:\ 1/8,\ n=46:\ 1/5$.

We also explored a variant of ExcLQA incorporating an enhanced formulation of the penalization approach of Ref.~\cite{julia}. While this alternative maintains the core methodology, it requires two penalization hyperparameters that need to be tuned via grid search. The results obtained using this variant are presented in App.~\ref{app:exclqa_exp}.

For further details on the hyperparameters used and the exact configuration of our simulations, refer to App.~\ref{app:hyperparam}.

\section{Conclusion \label{sec:conc}}
In this work, we have introduced ExcLQA, a physics-inspired algorithm for excited states of classical Hamiltonians. ExcLQA mimics a quantum adiabatic evolution from an initial Hamiltonian with a known ground state to a final cost function tailored to the low-energy configurations of the target Hamiltonian. This final cost function uses only one penalization hyperparameter, which can be tuned via binary search, to select the desired penalization level. The adiabatic evolution is restricted to pure product states, which avoids scalability issues on classical computers. 

To empirically demonstrate the effectiveness of ExcLQA, we benchmark it using instances of all-to-all interacting Ising models with random interactions and the shortest vector problem (SVP). For the random Ising models, we showed that ExcLQA outperforms state-of-the-art solvers based on simulated annealing and tensor-network methods, which require knowledge of the exact ground-state configuration, achieving a solved ratio of $0.87$ for the first excited state when the exact ground-state energy is provided. By appropriately tuning the penalty hyperparameter, ExcLQA can also target higher excited states, with solved ratios in the range $0.67$–$0.74$ for the second, third, and fourth excited states. In addition, ExcLQA does not require knowledge of the exact ground-state configuration or its exact energy to identify excited states. For example, when obtaining a lower bound for the ground state with semidefinite programming, ExcLQA still attains a solved ratio of $0.70$ for the first excited state, while the ratios of the first four excited states remain above $0.61$. For the SVP, for sublattice ranks between $10$ and $39$, the solved ratio of ExcLQA remains consistently above $0.675$ (\cref{fig:solved_ratio}), with less than 40 shots required, showing a roughly stable trend across lattice ranks (\cref{fig:shots}). Additionally, our algorithm successfully addressed some instances up to rank 39 of the approximate version of the SVP, $\gamma$-SVP, in the two-norm with $\gamma<\sqrt{2}$, an $\afunc{NP}$-hard problem. In the instances where the search space with a local dimension of two contained a solution, it achieved an approximation factor $\gamma$ below $1.185$ (\cref{fig:aprox_factor}).

We acknowledge a significant difficulty in tuning the hyperparameters to solve the SVP as the number of spins of the Hamiltonian grows. Although the penalty parameter $\alpha$ in \cref{eq:fin_cost} can be tuned by binary search, ExcLQA requires many other hyperparameters that require less efficient tuning strategies. We refer to App.~\ref{app:hyperparam} for details about the hyperparameters. This should not come as a surprise, as the Hamiltonians used to test ExcLQA correspond to the mapping of an $\afunc{NP}$-hard problem, which aligns with the observed increase in complexity as the sublattice rank grows. Note, however, that for the target Hamiltonian operators, ExcLQA offers a better performance than a Metropolis-Hastings optimization.

A natural extension of our work is to test ExcLQA on larger and more structured optimization problems where the ground state is either difficult to obtain or not necessarily the most useful solution in practice. Examples include scheduling, logistics, and resource-allocation tasks, where a simplified cost function may fail to capture secondary criteria such as robustness, fairness, or implementation constraints~\cite{Pinedo2016,Daskin2013}. A related example has recently appeared in machine-learning model compression, where transformer block removal was formulated as a constrained binary optimization problem and low-energy excited configurations were shown to outperform the ground state on downstream tasks~\cite{jansen_block_removal}. In such settings, the main appeal of ExcLQA is that it provides a direct way of searching beyond the ground state without requiring the ground-state configuration itself. It only requires a lower bound on the ground-state energy to shift the spectrum, and the excited-state penalty is controlled by a single parameter that can be tuned efficiently by binary search. This distinguishes ExcLQA from the approaches considered in Sec.~\ref{sec:fcIsing}, which require the ground-state configuration as an explicit input, either to penalize it in the QUBO objective or to enforce orthogonality to it~\cite{lucas2014ising,Stoudenmire_12}.

The present benchmarks demonstrate this mechanism on controlled instances, but whether the scalability previously observed for LQA in QUBO and integer-optimization problems~\cite{bowles,jansen_24} carries over to large-scale, controlled, excited-state search remains an important question for future work. More broadly, since the penalty term is defined at the level of the cost function, it could also be incorporated into other physics-inspired optimization methods, such as graph neural networks~\cite{schuetz_22, schuetz_22_2}. Finally, ExcLQA is based on a product-state ansatz and therefore does not exploit entanglement. This restriction makes the method classically efficient and allows for a more flexible choice of optimization objective, but further comparisons with more entanglement-aware methods~\cite{orus_14} could help clarify whether entanglement can provide an advantage for excited-state optimization.
\vspace{-3mm}
\section*{Code availability}
The implementation of our algorithm and the data used to generate the results presented in this work are publicly available in Ref.~\cite{exclqa_code}.
\vspace{-3mm}
\begin{acknowledgments}
We thank Anand Kumar Narayanan for suggestions on an earlier version of the manuscript. This work was supported by the ERC AdG CERQUTE, the AXA Chair in Quantum Information Science, the Government of Spain (Severo Ochoa CEX2019-000910-S, FUNQIP, European Union NextGenerationEU PRTR-C17.I1 and Quantum in Spain), the EU projects Veriqtas, QEC4QEA and PASQUANS2, Fundació Cellex, Fundació Mir-Puig, and Generalitat de Catalunya (CERCA program). JBR has received funding from the “Secretaria d’Universitats i Recerca del Departament de Recerca i Universitats de la Generalitat de Catalunya” under grant FI-2 00096, as well as the European Social Fund Plus.
\end{acknowledgments}
\vspace{-3mm}
\appendix
\label{ap}
\section{Binary-encoded qudits}\label{app:qudits-operators}
The problem described in \cref{eq:norm} seeks to find the integer coefficients $ \{x_i\}_{i=1}^n$ that minimize the squared length of a non-zero lattice vector. To enable a quantum emulation of this problem, one must construct a quantum Hamiltonian, such as the one proposed in \cref{eq:ham_svp}, where the operators $\hat{Q}^{(i)}$, when applied to a string of qubits, return the integers corresponding to the vector coefficients $\{x_i\}_{i=1}^n$. Since each qubit yields a measurement outcome of either 0 or 1, integers can be represented in binary across multiple qubits, with each bit determined by the measured state of an individual qubit.

Although these integer values are unbounded at first, for a practical implementation, we must assume they can be represented as binary numbers using $k$ bits. Thus, each coefficient $x_i$ is represented by a vector $(x_i^1, \dots, x_i^k)\in~\{0,1\}^k$. The action of the operators $\hat{Q}^{(i)}$ on a string of qubits representing the coefficients can be summarized as
\begin{equation}\label{eq:qudit_bin}
    \hat{Q}^{(i)} \ket{x_i^1 \dots x_i^k} = x_i \cdot \ket{x_i^1 \dots x_i^k}.
\end{equation}

In Ref.~\cite{joseph_map}, the authors introduce binary-encoded qudits, which provide a space-optimally efficient definition for the $\hat{Q}^{(i)}$ operators. Assuming each qudit is decomposed into $k$ qubits, the qudit operator acting on the $i$-th qudit can be expressed as
\begin{equation}
    \hat{Q}^{(i)} = \sum_{l=0}^{k-1} 2^l \hat{O}^{l,i} - 2^{k-1} \hat{\mathds{1}},
\end{equation}
where $\hat{O}^{l,i}$ is the $\hat{O} = (\hat{\mathds{1}} - \hat{\s}_z)/2$ operator acting on the $l$-th qubit of the $i$-th qudit. The first term in \cref{eq:qudit_bin} converts the binary string into an integer by interpreting it as a binary number, while the second term shifts the range to $[-2^{k-1}, 2^{k-1} - 1] \cap \mathbb{Z}$, which is approximately symmetric around zero, and allows for negative coefficients. This method achieves optimal space efficiency as the mapping between the spin configurations and the vector coefficients is injective.

\vspace{-3mm}
\section{Metropolis-Hastings algorithm}\label{app:metropolis}
In this section, we provide a brief overview of the Metropolis-Hastings algorithm~\cite{metropolis1953equation, metropolis} following Ref.~\cite[Section 7.3]{robert2004metropolis}. 

The Metropolis-Hastings algorithm is a Markov Chain Monte Carlo method used to sample from a target probability distribution, $\pi(x)$, when direct sampling is impractical. It constructs a Markov chain, i.e., a sequence of samples $\{x^{(0)}, x^{(1)}, \dots, x^{(N)}\}$ where each $x^{(i+1)}$ only depends on the previous one, $x^{(i)}$. The process starts with an initial sample $x^{(0)}$, which can be chosen arbitrarily, though a better initialization can accelerate convergence. At each step, a new candidate state $y_i$ is proposed from a distribution $q(y_i|x^{(i)})$, defining how the Markov chain moves through the state space. The candidate is then accepted or rejected based on an acceptance probability $\rho(x^{(i)},y_i)$ that guides the chain's exploration. By iterating this process, the algorithm efficiently navigates the state space, progressively generating samples that approximate $\pi(x)$, meaning that as $N \to \infty$, the proportion of visits to each state equals its probability under $\pi(x)$.

Beyond probabilistic sampling, the Metropolis-Hastings framework can be adapted for optimization by reformulating the search for a minimum as a stochastic process over the solution space. In our case, we optimize the cost function $E_{\text{F}}$ defined in \cref{eq:fin_cost}, which does not represent the energy of a physical Hamiltonian but is instead designed to penalize undesired global minima.

To apply the Metropolis algorithm in this setting, we define a target distribution as

\begin{equation}
    \pi(x) \propto \exp\left(-\frac{E_{\text{F}}(x)}{T}\right),
\end{equation}
where $T$ is a temperature parameter that controls the balance between exploration and exploitation. At high temperatures, the distribution $\pi(x)$ is broader, increasing the acceptance of higher-cost states and promoting exploration. As $T$ decreases, the distribution concentrates around low-cost states, favoring exploitation and guiding the algorithm toward an optimal solution.

The implemented algorithm follows a standard Metropolis scheme, where, at each iteration, a new configuration $y_i$ is proposed by flipping a single randomly chosen spin in the current configuration $x^{(i)}$. The acceptance probability for this proposed configuration is given by

\begin{equation}
    \rho(x^{(i)}, y_i) = \min\left\{1, \exp\left(-\frac{E_{\text{F}}(y_i) - E_{\text{F}}(x^{(i)})}{T} \right) \right\}.
\end{equation}
If the proposed configuration $y_i$ is accepted, the system state is updated to $x^{(i+1)} = y_i$; otherwise, the current configuration is retained. We summarize the method in Alg.~\ref{alg_mh}.

\vspace{2mm}
\begin{algorithm}[H]\label{alg_mh}
\KwIn{$E_{\text{F}}(x)$: cost function; $N$: total iterations; $T$: temperature.}
\KwOut{$x^*$: optimized configuration.}
$x^{(0)}\gets$ random initial spin configuration\\
$x^* \gets x^{(0)}$\\
\For{$i=1,\dots, N$}
{
    Select random spin index $j$\\
    $y_i \gets x^{(i-1)}$ with flipped spin $j$\\
\resizebox{.9\linewidth}{!}{$
\rho(x^{(i-1)},y_i) \gets \min\left\{1, \exp\left(-\frac{E_{\text{F}}(y_i) - E_{\text{F}}(x^{(i-1)})}{T} \right)\right\}
$}\\
    $x^{(i)}\gets 
    \begin{cases} 
        y_i & \text{with prob. } \rho(x^{(i-1)},y_i), \\
        x^{(i-1)} & \text{with prob. } 1-\rho(x^{(i-1)},y_i).
    \end{cases}$\\
    \If{$E_{\text{F}}(x^{(i)}) < E_{\text{F}}(x^*)$}
    {
        $x^* \gets x^{(i)}$
    }
}
\Return{$x^*$}
\caption{Metropolis optimization}
\end{algorithm}
\vspace{-2mm}

\section{ExcLQA with alternative cost}\label{app:exclqa_exp}
In this appendix, we introduce an alternative final cost function, motivated by the approach in Ref.~\cite{julia}, defined as \begin{equation}\label{eq:cost-julia}
E_\text{F}(\boldsymbol{\theta})\coloneqq \expval{\smash{\hat{H}_z}}+re^{-s\expval{\smash{\hat{H}_z}}}, 
\end{equation} where $r,s\in\R_{>0}$ are tunable hyperparameters. The term $re^{-s\expval{\smash{\hat{H}_z}}}$ raises the energy of the original ground state to $r$, while remaining exponentially small for excited states. Note that this formulation also requires the spectrum of $\hat{H}_z$ to be nonnegative to prevent divergence of the exponential term. If needed, this can be ensured by shifting the Hamiltonian with a constant offset, as discussed in the main text.

\begin{figure}[ht]
    \centering
    \begin{subfigure}[a]{1\columnwidth}
        \includegraphics[width=\textwidth]{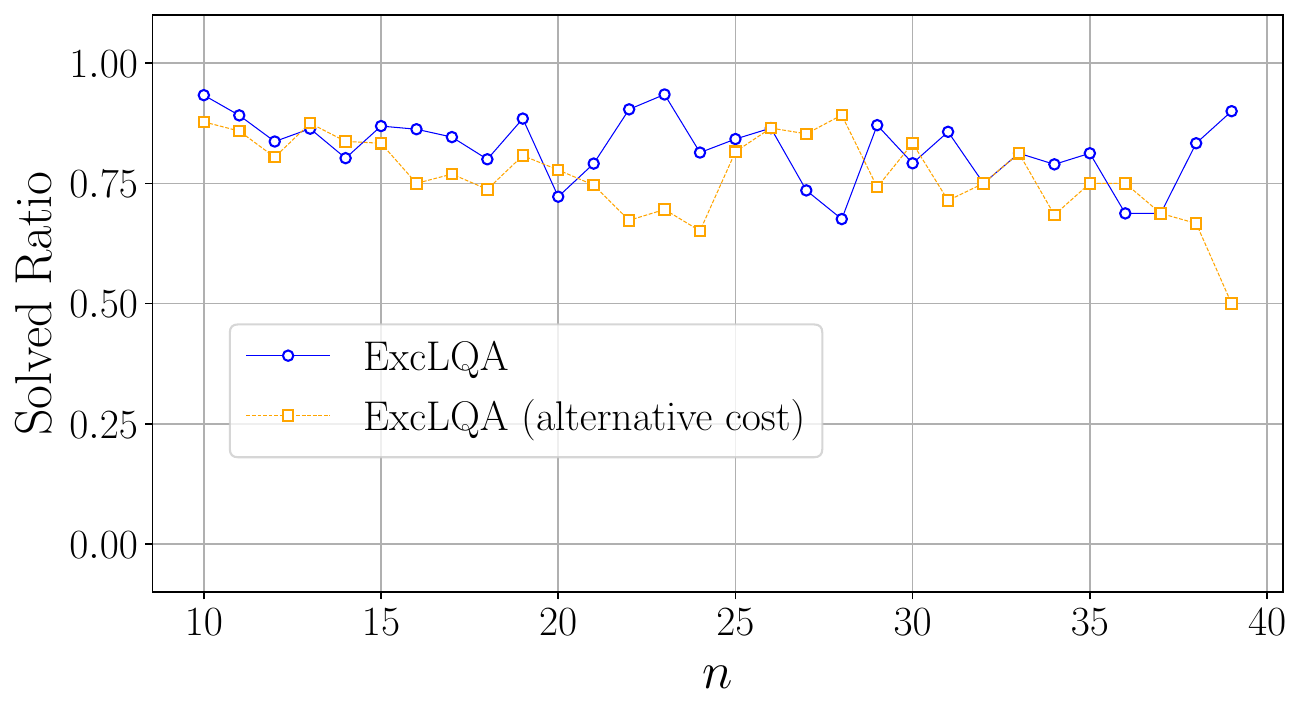}
        \caption{Success rate of instances solved using each method.}
        \label{fig:solved_ratio_2}
    \end{subfigure}
    \begin{subfigure}[b]{1\columnwidth}
        \includegraphics[width=\textwidth]{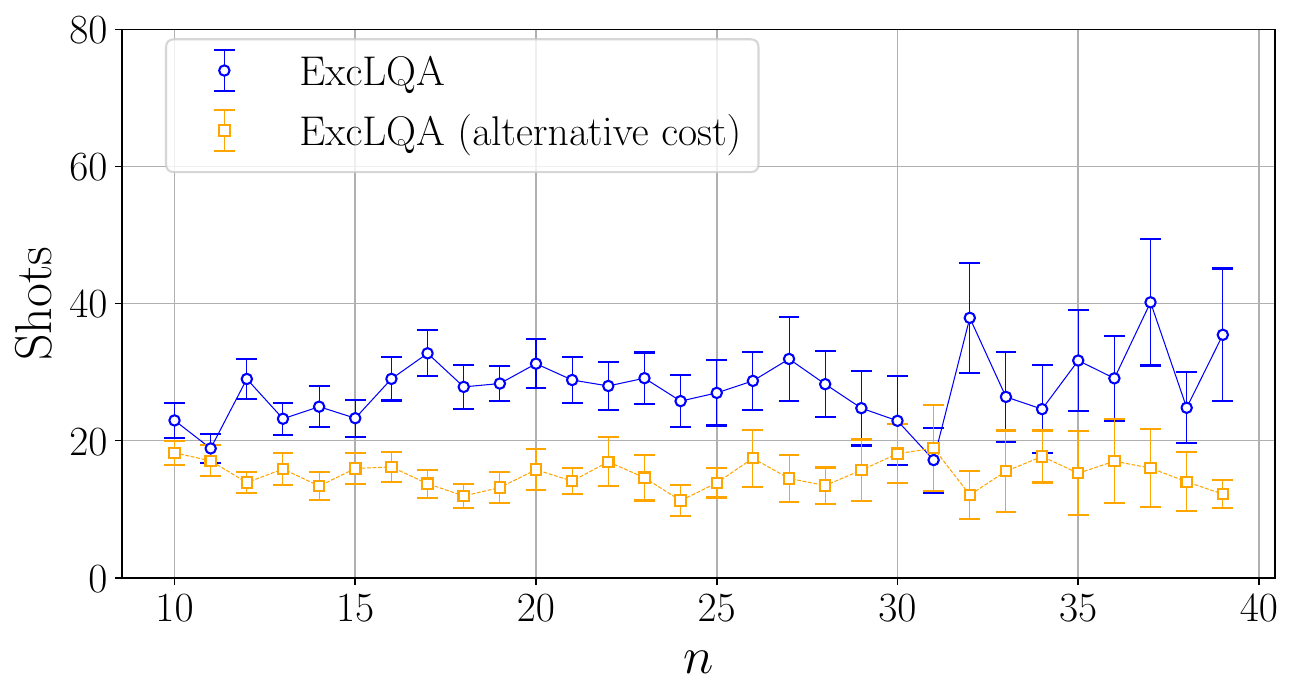}
        \caption{Average number of shots required to find a solution.}
        \label{fig:shots_2}
    \end{subfigure}
    \begin{subfigure}[c]{1\columnwidth}
        \includegraphics[width=\textwidth]{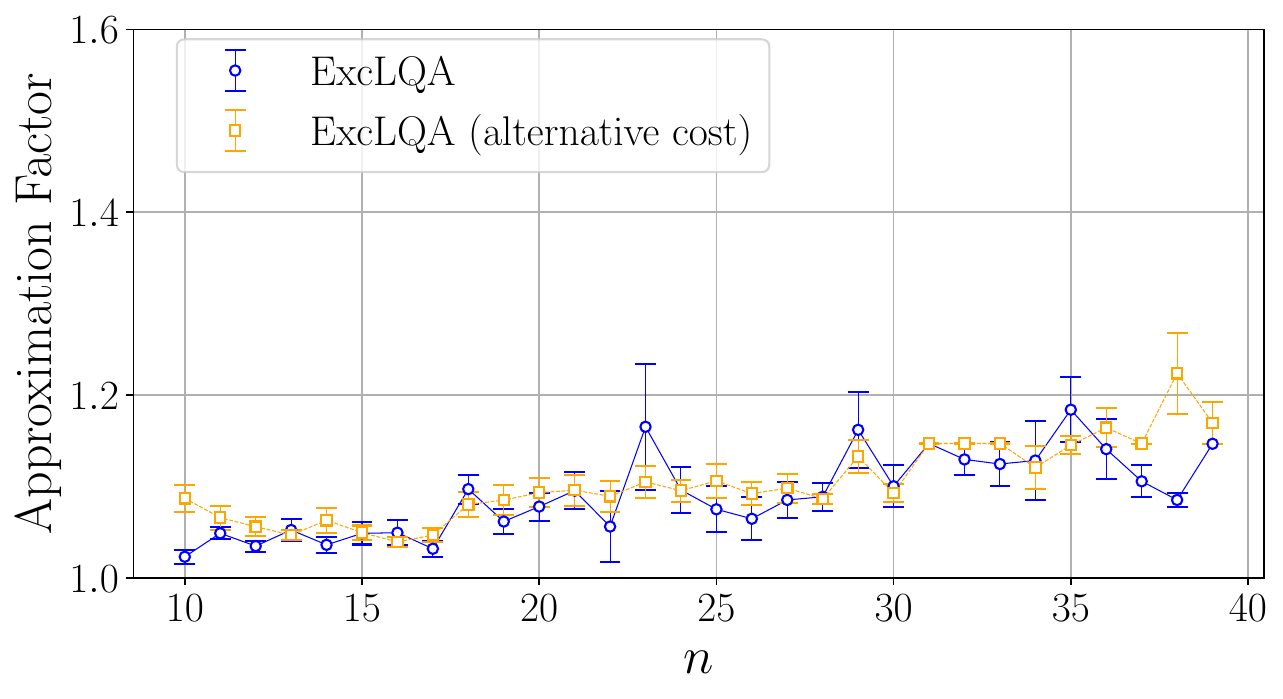}
        \caption{Average approximation factor for unsolved instances.}
        \label{fig:aprox_factor_2}
    \end{subfigure}
    \caption{\justifying Summary of performance metrics for ExcLQA and its alternative cost function. We generated $100$ problem instances for each sublattice rank $n$ and restricted the analysis to those whose solutions were contained within the search space defined in \cref{eq:search_space_k} with a local dimension of 2.}
    \label{fig:results_2}
\end{figure}

Fig.~\ref{fig:results_2} summarizes the results obtained for both ExcLQA and the alternative cost function. The solved ratio as a function of the sublattice rank $n$ is shown in Fig.~\ref{fig:solved_ratio_2}. While both methods perform similarly for small $n$, the solved ratio for the alternative cost function decreases more rapidly as $n$ increases. Notably, for $n \geq 36$, its solved ratio drops below 0.75 and reaches 0.5 at $n=39$, whereas the main ExcLQA method maintains a higher solved ratio across most ranks.

Fig.~\ref{fig:shots_2} displays the average number of shots required to find a solution. The alternative cost function consistently requires fewer shots than the main ExcLQA method. This is likely because the hyperparameters identified for the alternative cost function required 
$N=4000$ points during the adiabatic evolution, providing greater stability at the cost of significantly increased computational time. In contrast, ExcLQA achieved better performance with only $N=100$ points, making it considerably more efficient.

Finally, Fig.~\ref{fig:aprox_factor_2} presents the approximation factor $\gamma$ for instances where the SVP was not solved within 100 shots. While both methods maintain relatively low approximation factors, the alternative cost function exhibits a steeper increase in $\gamma$ for larger $n$. This suggests that when the solver fails to find an optimal solution, ExcLQA provides more reliable approximations compared to the alternative method.

By restricting ExcLQA with the alternative cost function to the search space defined in \cref{eq:search_space_k} with a local dimension of 4, where each vector coefficient is either $-2, -1, 0,$ or $1$, we extended its applicability to higher ranks. However, this significantly increased the difficulty of hyperparameter tuning. The following fractions represent the number of successfully solved instances over the total considered within the search space: $n=43:\ 2/16,\ n=44:\ 2/12,\ n=45:\ 1/8,\ n=46:\ 1/5$. For a detailed explanation of the hyperparameters used, we refer to App.~\ref{app:hyperparam}.

\vspace{-3mm}
\section{Hyperparameters}\label{app:hyperparam}
The hyperparameters required for ExcLQA and its alternative formulation are detailed in Tab.~\ref{table:param_expl}. For ExcLQA, the penalty parameter $\alpha$ was selected via a binary search procedure, while the remaining hyperparameters were manually tuned. For the alternative cost function, a grid search was employed to tune $r$ and $s$ in \cref{eq:cost-julia}. Specifically, tuning $r$ required an estimator for the squared length of the shortest non-zero vector, for which we used the square of the Gaussian heuristic. A grid search was then performed to determine the constant $A$ in $r = A\cdot\gh(\Lat)^2$, while $s$ was selected via a grid search with fixed precision.

We list below the hyperparameters used in the illustrative simulations shown in Figs.~\ref{fig:penalising} and~\ref{fig:energy}.
\begin{itemize}
    \item Fig.~\ref{fig:penalising}: $N=100$, $f=0.2$, $\mu=0.999$, $\eta=0.999$, $\kappa=6$, and $\alpha=0.55$.
    \item Fig.~\ref{fig:energy}: $N=100$, $f=0.2$, $\mu=0.999$, $\eta=0.999$, $\kappa=8$, and $\alpha=0.055$. In addition, the Gram matrix is rescaled as $\vec{G}/\norm{\vec{G}}$.
\end{itemize}

In Tab.~\ref{table:performance_exclqa} we detail the results of our simulations in Sec.~\ref{sec:results} for ranks ranging from 10 to 39. We call valid instances those SVP instances whose solution was contained in our search space.

\newpage
\onecolumngrid
\begin{center}
\begin{table}[htbp]
\renewcommand{\arraystretch}{1.4}
\begin{tabular}{ |c|c|c|c|c|c|} 
 \hline
Hyperparameter & Description & \multicolumn{4}{c|}{Value} \\ 
 \cline{3-6}
 &  & \multicolumn{2}{c|}{ExcLQA} & \multicolumn{2}{c|}{Alternative} \\  
  \cline{3-6}
  &  & \multicolumn{2}{c|}{Local Dimension} & \multicolumn{2}{c|}{Local Dimension} \\  
  \cline{3-6}
  &  & 2 & 4 & 2 & 4 \\  
 \hline
 $N$  & Points in the interval $[0,1]$ to simulate an adiabatic evolution. & 100 & 250 & 4000 & 100  \\ 
 $\kappa$ & Strength of the function $E_{\text{F}}$ in \cref{eq:total_cost}. & 8 & 30 & 1 & $0.008$ \\ 
 $\mu$ & Momentum in the SGD optimizer. & $0.9989$ & $0.999$ & $0.9989$ & $0.999999$\\
 $\eta$ & Learning rate in the SGD optimizer. & $0.999$ & $0.009$ & $0.0091$ & $0.091$ \\ 
 $M$ & Rescaling factor for the Gram matrix in \cref{eq:ham_svp}. & $\norm{\vec{G}}$ & $\norm{\vec{G}}/50$ & 1 & 16385\\
 $\alpha$ & Prefactor in $1/\expval{\smash{\hat{H}_z}}$ from \cref{eq:fin_cost}. & $0.055$ & $3\cdot 10^{-10}$ & --- & ---\\
 $s$ & Prefactor in $-\expval{\smash{\hat{H}_z}}$ from \cref{eq:cost-julia}. & --- & --- & $4.6\cdot 10^{-7}$ & $0.0005$\\
 $r$ & Prefactor in $\exp\left(-s\expval{\smash{\hat{H}_z}}\right)$ from \cref{eq:cost-julia}. & --- & --- & $0.72\cdot \gh(\Lat)^2$ & $0.72\cdot \gh(\Lat)^2$\\
 $f$ & Sampling interval $[-f,f]$ for each qubit in the initial state. & $0.2$ & $0.2$ & $0.15$ & $0.15$\\
 \hline
\end{tabular}
\caption{Summary of the hyperparameters for ExcLQA and its alternative cost function.}
\label{table:param_expl}
\end{table}
\end{center}

\vspace{-10mm}
\begin{center}
\begin{table}[!htbp]
\small
\renewcommand{\arraystretch}{1.22}
\begin{tabular}{ |c|c|c|c|c|c|c|c| } 
 \hline
 $n$ & Valid Instances & \multicolumn{2}{c|}{Solved Instances} & \multicolumn{2}{c|}{Avg. Shot Number} & \multicolumn{2}{c|}{Avg. Approx. Factor} \\ 
 \cline{3-8}
  &  & ExcLQA & Alternative & ExcLQA & Alternative & ExcLQA & Alternative \\  
 \hline
 10 & 90 & 84 & 79 & 22.952 & 18.228 & 1.023 & 1.087 \\ 
 11 & 92 & 82 & 79 & 18.866 & 17.051 & 1.049 & 1.066 \\
 12 & 92 & 77 & 74 & 29.000 & 13.878 & 1.035 & 1.056 \\
 13 & 88 & 76 & 77 & 23.197 & 15.870 & 1.052 & 1.047 \\
 14 & 86 & 69 & 72 & 24.957 & 13.375 & 1.036 & 1.063 \\
 15 & 84 & 73 & 70 & 23.274 & 15.943 & 1.049 & 1.049 \\
 16 & 80 & 69 & 60 & 29.014 & 16.167 & 1.049 & 1.039 \\
 17 & 78 & 66 & 60 & 32.742 & 13.717 & 1.032 & 1.047 \\
 18 & 80 & 64 & 59 & 27.844 & 11.932 & 1.097 & 1.080 \\
 19 & 78 & 69 & 63 & 28.333 & 13.159 & 1.062 & 1.085 \\
 20 & 72 & 52 & 56 & 31.231 & 15.768 & 1.078 & 1.093 \\
 21 & 67 & 53 & 50 & 28.849 & 14.100 & 1.095 & 1.096 \\
 22 & 52 & 47 & 35 & 27.979 & 16.914 & 1.056 & 1.089 \\
 23 & 46 & 43 & 32 & 29.116 & 14.594 & 1.165 & 1.105 \\
 24 & 43 & 35 & 28 & 25.771 & 11.250 & 1.096 & 1.095 \\
 25 & 38 & 32 & 31 & 26.969 & 13.839 & 1.075 & 1.106 \\
 26 & 37 & 32 & 32 & 28.719 & 17.438 & 1.065 & 1.092 \\
 27 & 34 & 25 & 29 & 31.920 & 14.483 & 1.085 & 1.098 \\
 28 & 37 & 25 & 33 & 28.240 & 13.455 & 1.089 & 1.086 \\
 29 & 31 & 27 & 23 & 24.741 & 15.696 & 1.162 & 1.133 \\
 30 & 24 & 19 & 20 & 22.895 & 18.100 & 1.100 & 1.093 \\ 
 31 & 14 & 12 & 10 & 17.167 & 18.900 & 1.147 & 1.147 \\ 
 32 & 16 & 12 & 12 & 37.917 & 12.083 & 1.129 & 1.147 \\ 
 33 & 16 & 13 & 13 & 26.385 & 15.538 & 1.124 & 1.147 \\ 
 34 & 19 & 15 & 13 & 24.600 & 17.692 & 1.128 & 1.121 \\ 
 35 & 16 & 13 & 12 & 31.692 & 15.250 & 1.184 & 1.145 \\ 
 36 & 16 & 11 & 12 & 29.091 & 17.000 & 1.141 & 1.164 \\ 
 37 & 16 & 11 & 11 & 40.182 & 16.000 & 1.106 & 1.147 \\ 
 38 & 12 & 10 & 8  & 24.800 & 14.000 & 1.085 & 1.224 \\ 
 39 & 10 & 9  & 5  & 35.444 & 12.200 & 1.147 & 1.169 \\ 
 \hline
\end{tabular}
\caption{Detailed results from ExcLQA and its alternative cost function for sublattice ranks ranging from 10 to 39 with a local dimension of 2.}
\label{table:performance_exclqa}
\end{table}
\end{center}
\clearpage
\twocolumngrid

\newpage
\bibliographystyle{biblev1}
\bibliography{bib}

@article{LLL,
  title={Factoring polynomials with rational coefficients},
  author={Lenstra, Arjen K and Lenstra, Hendrik Willem and Lov{\'a}sz, L{\'a}szl{\'o}},
  journal={Mathematische annalen},
  volume={261},
  pages={515--534},
  year={1982},
  publisher={Springer Verlag}
}

@book{minkowskith,
  title={Geometrie der zahlen},
  author={Minkowski, Hermann},
  volume={1},
  year={1910},
  publisher={BG Teubner}
}

@article{bowles,
  title = {Quadratic Unconstrained Binary Optimization via Quantum-Inspired Annealing},
  author = {Bowles, Joseph and Dauphin, Alexandre and Huembeli, Patrick and Martinez, Jos\'e and Ac\'{\i}n, Antonio},
  journal = {Phys. Rev. Appl.},
  volume = {18},
  issue = {3},
  pages = {034016},
  numpages = {10},
  year = {2022},
  month = {Sep},
  publisher = {American Physical Society},
  doi = {10.1103/PhysRevApplied.18.034016},
  url = {https://link.aps.org/doi/10.1103/PhysRevApplied.18.034016}
}

@article{joseph_map,
  title = {Two quantum {Ising} algorithms for the shortest-vector problem},
  author = {Joseph, David and Callison, Adam and Ling, Cong and Mintert, Florian},
  journal = {Phys. Rev. A},
  volume = {103},
  issue = {3},
  pages = {032433},
  numpages = {12},
  year = {2021},
  month = {Mar},
  publisher = {American Physical Society},
  doi = {10.1103/PhysRevA.103.032433},
  url = {https://link.aps.org/doi/10.1103/PhysRevA.103.032433}
}

@article{not_so_adiabatic,
  title = {Not-so-adiabatic quantum computation for the shortest vector problem},
  author = {Joseph, David and Ghionis, Alexandros and Ling, Cong and Mintert, Florian},
  journal = {Phys. Rev. Res.},
  volume = {2},
  issue = {1},
  pages = {013361},
  numpages = {13},
  year = {2020},
  month = {Mar},
  publisher = {American Physical Society},
  doi = {10.1103/PhysRevResearch.2.013361},
  url = {https://link.aps.org/doi/10.1103/PhysRevResearch.2.013361}
}

@article{julia,
  title = {Finding dense sublattices as low energy states of a {Hamiltonian}},
  author = {Barber\`a-Rodr\'{\i}guez, J\'ulia and Gama, Nicolas and Narayanan, Anand Kumar and Joseph, David},
  journal = {Phys. Rev. Res.},
  volume = {6},
  issue = {4},
  pages = {043279},
  numpages = {18},
  year = {2024},
  month = {Dec},
  publisher = {American Physical Society},
  doi = {10.1103/PhysRevResearch.6.043279},
  url = {https://link.aps.org/doi/10.1103/PhysRevResearch.6.043279}
}

@article{Albrecht_2023,
   title={Variational quantum solutions to the Shortest Vector Problem},
   volume={7},
   ISSN={2521-327X},
   url={http://dx.doi.org/10.22331/q-2023-03-02-933},
   DOI={10.22331/q-2023-03-02-933},
   journal={Quantum},
   publisher={Verein zur Forderung des Open Access Publizierens in den Quantenwissenschaften},
   author={Albrecht, Martin R. and Prokop, Miloš and Shen, Yixin and Wallden, Petros},
   year={2023},
   month=mar, pages={933} }

@inproceedings{red_lat,
author = {Ajtai, M.},
title = {Generating hard instances of lattice problems (extended abstract)},
year = {1996},
isbn = {0897917855},
publisher = {Association for Computing Machinery},
address = {New York, NY, USA},
url = {https://doi.org/10.1145/237814.237838},
doi = {10.1145/237814.237838},
booktitle = {Proceedings of the Twenty-Eighth Annual ACM Symposium on Theory of Computing},
pages = {99-–108},
numpages = {10},
location = {Philadelphia, Pennsylvania, USA},
series = {STOC '96}
}

@Article{Micciancio01svp,
  author =  {Daniele Micciancio},
  title =  {The Shortest Vector Problem is {NP}-hard to approximate to within some constant},
  journal =  {SIAM Journal on Computing},
  year =  2001,
  volume = 30,
  number = 6,
  pages = {2008--2035},
  month = mar,
  note = {Preliminary version in FOCS 1998},
  doi  = {10.1137/S0097539700373039}
}

@InProceedings{complexity_svp,
author="Ducas, L{\'e}o
and Stevens, Marc
and van Woerden, Wessel",
editor="Canteaut, Anne
and Standaert, Fran{\c{c}}ois-Xavier",
title="Advanced Lattice Sieving on {GPUs}, with Tensor Cores",
booktitle="Advances in Cryptology -- EUROCRYPT 2021",
year="2021",
publisher="Springer International Publishing",
address="Cham",
pages="249--279",
isbn="978-3-030-77886-6"
}

@article{SGD,
  title={A stochastic approximation method},
  author={Robbins, Herbert and Monro, Sutton},
  journal={The annals of mathematical statistics},
  pages={400--407},
  year={1951},
  publisher={JSTOR}
}

@article{pytorch,
  title={Pytorch: An imperative style, high-performance deep learning library},
  author={Paszke, Adam and Gross, Sam and Massa, Francisco and Lerer, Adam and Bradbury, James and Chanan, Gregory and Killeen, Trevor and Lin, Zeming and Gimelshein, Natalia and Antiga, Luca and others},
  journal={Advances in neural information processing systems},
  volume={32},
  year={2019}
}

@article{lwe,
author = {Regev, Oded},
title = {On Lattices, Learning with Errors, Random Linear Codes, and Cryptography},
year = {2009},
issue_date = {September 2009},
publisher = {Association for Computing Machinery},
address = {New York, NY, USA},
volume = {56},
number = {6},
issn = {0004-5411},
url = {https://doi.org/10.1145/1568318.1568324},
doi = {10.1145/1568318.1568324},
journal = {J. ACM},
month = {sep},
articleno = {34},
numpages = {40}
}

@article{bose_hubbard,
  title = {Not-so-adiabatic quantum computation for the shortest vector problem},
  author = {Joseph, David and Ghionis, Alexandros and Ling, Cong and Mintert, Florian},
  journal = {Phys. Rev. Res.},
  volume = {2},
  issue = {1},
  pages = {013361},
  numpages = {13},
  year = {2020},
  month = {Mar},
  publisher = {American Physical Society},
  doi = {10.1103/PhysRevResearch.2.013361},
  url = {https://link.aps.org/doi/10.1103/PhysRevResearch.2.013361}
}

@article{jansen_24,
  title = {Qudit-inspired optimization for graph coloring},
  author = {Jansen, David and Heightman, Timothy and Mortimer, Luke and Perito, Ignacio and Ac\'{\i}n, Antonio},
  journal = {Phys. Rev. Appl.},
  volume = {22},
  issue = {6},
  pages = {064002},
  numpages = {12},
  year = {2024},
  month = {Dec},
  publisher = {American Physical Society},
  doi = {10.1103/PhysRevApplied.22.064002},
  url = {https://link.aps.org/doi/10.1103/PhysRevApplied.22.064002}
}

@article{tesoro_24,
  title = {Integer factorization via tensor-network Schnorr's sieving},
  author = {Tesoro, Marco and Siloi, Ilaria and Jaschke, Daniel and Magnifico, Giuseppe and Montangero, Simone},
  journal = {Phys. Rev. A},
  volume = {113},
  issue = {3},
  pages = {032418},
  numpages = {15},
  year = {2026},
  month = {Mar},
  publisher = {American Physical Society},
  doi = {10.1103/d9dl-ctt4},
  url = {https://link.aps.org/doi/10.1103/d9dl-ctt4}
}

@article{veszeli_21,
  title={Mean field approximation for solving QUBO problems},
  author={Veszeli, M{\'a}t{\'e} Tibor and Vattay, G{\'a}bor},
  journal={Plos one},
  volume={17},
  number={8},
  pages={e0273709},
  year={2022},
  publisher={Public Library of Science San Francisco, CA USA}
}

@Misc{fioroni_25,
      title={Entanglement-assisted variational algorithm for discrete optimization problems}, 
      author={  Fioroni, L. and  Savona, V.},
      year={2025},
      HowPublished={arXiv:2501.09078},
      Doi={arXiv:2501.09078},
      url={https://doi.org/10.48550/arXiv.2501.09078}, 
}

@inproceedings{Peikert_red,
  title={Public-key cryptosystems from the worst-case shortest vector problem},
  author={Peikert, Chris},
  booktitle={Proceedings of the forty-first annual ACM symposium on Theory of computing},
  pages={333--342},
  year={2009}
}

@article{schuetz_22,
  title = {Combinatorial optimization with physics-inspired graph neural networks},
  author = {Schuetz, Martin J. A. and Brubaker, J. Kyle and Katzgraber, Helmut G.},
  journal = {Nat. Mach. Intell.},
  volume = {4},
  issue = {4},
  pages = {367–377},
  numpages = {10},
  year = {2022},
  doi = {10.1038/s42256-022-00468-6},
  url = {https://doi.org/10.1038/s42256-022-00468-6}
}

@article{schuetz_22_2,
  title = {Graph coloring with physics-inspired graph neural networks},
  author = {Schuetz, Martin J. A. and Brubaker, J. Kyle and Zhu, Zhihuai and Katzgraber, Helmut G.},
  journal = {Phys. Rev. Res.},
  volume = {4},
  issue = {4},
  pages = {043131},
  numpages = {10},
  year = {2022},
  month = {Nov},
  publisher = {American Physical Society},
  doi = {10.1103/PhysRevResearch.4.043131},
  url = {https://link.aps.org/doi/10.1103/PhysRevResearch.4.043131}
}

@unpublished{fplll,
    author = {The {FPLLL} development team},
    title = {{fplll}, a lattice reduction library, {Version}: 5.4.5},
    year = 2023,
    note = {Available at \url{https://github.com/fplll/fplll}},
    url = {https://github.com/fplll/fplll}
}

@article{orus_14,
title = {A practical introduction to tensor networks: Matrix product states and projected entangled pair states},
journal = {Annals of Physics},
volume = {349},
pages = {117-158},
year = {2014},
issn = {0003-4916},
doi = {https://doi.org/10.1016/j.aop.2014.06.013},
url = {https://www.sciencedirect.com/science/article/pii/S0003491614001596},
author = {Román Orús}
}

@article{metropolis,
    author = {Hastings, W. K.},
    title = {Monte Carlo sampling methods using Markov chains and their applications},
    journal = {Biometrika},
    volume = {57},
    number = {1},
    pages = {97-109},
    year = {1970},
    month = {04},
    issn = {0006-3444},
    doi = {10.1093/biomet/57.1.97},
    url = {https://doi.org/10.1093/biomet/57.1.97},
}

@article{metropolis1953equation,
  title={Equation of state calculations by fast computing machines},
  author={Metropolis, Nicholas and Rosenbluth, Arianna W and Rosenbluth, Marshall N and Teller, Augusta H and Teller, Edward},
  journal={The journal of chemical physics},
  volume={21},
  number={6},
  pages={1087--1092},
  year={1953},
  publisher={American Institute of Physics}
}

@article{robert2004metropolis,
  title={The metropolis—hastings algorithm},
  author={Robert, Christian P and Casella, George and Robert, Christian P and Casella, George},
  journal={Monte Carlo statistical methods},
  pages={267--320},
  year={2004},
  publisher={Springer}
}

@Misc{Farhi2014,
  author    = {Edward Farhi and Jeffrey Goldstone and Sam Gutmann},
  title     = {A quantum approximate optimization algorithm},
  journal   = {arXiv preprint},
  year      = {2014},
  HowPublished = {arXiv:1411.4028},
  Doi = {arXiv:1411.4028},
  url = {https://doi.org/10.48550/arXiv.1411.4028}
}

@book{Wolsey1998,
  author    = {Laurence A. Wolsey},
  title     = {Integer Programming},
  publisher = {Wiley},
  year      = {1998}
}

@book{Daskin2013,
  author    = {Mark S. Daskin},
  title     = {Network and Discrete Location: Models, Algorithms, and Applications},
  edition   = {2nd},
  publisher = {Wiley},
  year      = {2013}
}

@book{Pinedo2016,
  author    = {Michael L. Pinedo},
  title     = {Scheduling: Theory, Algorithms, and Systems},
  edition   = {5th},
  publisher = {Springer},
  year      = {2016}
}

@ARTICLE{prokop2025heuristictimecomplexitynisq,
  author={Prokop, Miloš and Wallden, Petros},
  journal={IEEE Transactions on Quantum Engineering}, 
  title={Heuristic Time Complexity of NISQ Shortest-Vector-Problem Solvers}, 
  year={2025},
  volume={6},
  number={},
  pages={1-19},
  doi={10.1109/TQE.2025.3620104}}

@misc{glover_18,
      title={A Tutorial on Formulating and Using {QUBO} Models}, 
      author={Glover, F. and  Kochenberger, G. and  Du, Y.},
      year={2025},
      HowPublished={arXiv:1811.11538},
      Doi={arXiv:1811.11538},
      url={https://doi.org/10.48550/arXiv.1811.1153} 
}

@article{mugel_22,
  title = {Dynamic portfolio optimization with real datasets using quantum processors and quantum-inspired tensor networks},
  author = {Mugel, Samuel and Kuchkovsky, Carlos and S\'anchez, Escol\'astico and Fern\'andez-Lorenzo, Samuel and Luis-Hita, Jorge and Lizaso, Enrique and Or\'us, Rom\'an},
  journal = {Phys. Rev. Res.},
  volume = {4},
  issue = {1},
  pages = {013006},
  numpages = {12},
  year = {2022},
  month = {Jan},
  publisher = {American Physical Society},
  doi = {10.1103/PhysRevResearch.4.013006},
  url = {https://link.aps.org/doi/10.1103/PhysRevResearch.4.013006}
}

@article{Kirkpatrick1983,
author = {S. Kirkpatrick  and C. D. Gelatt  and M. P. Vecchi },
title = {Optimization by Simulated Annealing},
journal = {Science},
volume = {220},
number = {4598},
pages = {671-680},
year = {1983},
doi = {10.1126/science.220.4598.671},
url = {https://www.science.org/doi/abs/10.1126/science.220.4598.671}
}

@article{Hukushima1996,
  title={Exchange Monte Carlo method and application to spin glass simulations},
  author={Hukushima, Koji and Nemoto, Koji},
  journal={Journal of the Physical Society of Japan},
  volume={65},
  number={6},
  pages={1604--1608},
  year={1996},
  publisher={The Physical Society of Japan}
}

@article{goemans1995improved,
  title={Improved approximation algorithms for maximum cut and satisfiability problems using semidefinite programming},
  author={Goemans, Michel X and Williamson, David P},
  journal={Journal of the ACM (JACM)},
  volume={42},
  number={6},
  pages={1115--1145},
  year={1995},
  publisher={ACM New York, NY, USA}
}

@article{Anderson1951,
  title = {Limits on the Energy of the Antiferromagnetic Ground State},
  author = {Anderson, P. W.},
  journal = {Phys. Rev.},
  volume = {83},
  issue = {6},
  pages = {1260--1260},
  numpages = {0},
  year = {1951},
  month = {Sep},
  publisher = {American Physical Society},
  doi = {10.1103/PhysRev.83.1260},
  url = {https://link.aps.org/doi/10.1103/PhysRev.83.1260}
}

@book{Papadimitriou1982,
  title={Combinatorial Optimization: Algorithms and Complexity},
  author={Papadimitriou, Christos H and Steiglitz, Kenneth},
  publisher={Prentice-Hall},
  year={1982}
}

@book{Baxter1982,
  title={Exactly Solved Models in Statistical Mechanics},
  author={Baxter, Rodney J},
  publisher={Academic Press},
  year={1982},
  address={London}
}

@Inbook{Glover1998,
author="Glover, Fred and Laguna, Manuel",
editor="Du, Ding-Zhu and Pardalos, Panos M.",
title="Tabu Search",
bookTitle="Handbook of Combinatorial Optimization: Volume1--3",
year="1998",
publisher="Springer US",
address="Boston, MA",
pages="2093--2229",
isbn="978-1-4613-0303-9",
doi="10.1007/978-1-4613-0303-9_33",
url="https://doi.org/10.1007/978-1-4613-0303-9_33"
}

@article{lucas2014ising,
  title={Ising formulations of many {NP} problems},
  author={Lucas, Andrew},
  journal={Frontiers in physics},
  volume={2},
  pages={5},
  year={2014},
  publisher={Frontiers Media SA}
}

@article{Oganov2006,
  title={Crystal structure prediction using ab initio evolutionary techniques: Principles and applications},
  author={Oganov, Artem R and Glass, Colin W},
  journal={The Journal of chemical physics},
  volume={124},
  number={24},
  year={2006},
  publisher={AIP Publishing}
}

@article{Anfinsen1973,
  title={Principles that govern the folding of protein chains},
  author={Anfinsen, Christian B},
  journal={Science},
  volume={181},
  number={4096},
  pages={223--230},
  year={1973},
  publisher={American Association for the Advancement of Science}
}

@book{baker2018principles,
  title={Principles of sequencing and scheduling},
  author={Baker, Kenneth R and Trietsch, Dan},
  year={2018},
  publisher={John Wiley \& Sons}
}

@article{Farhi2001,
  title = {Quantum Computation by Adiabatic Evolution},
  author = {Farhi, Edward and Goldstone, Jeffrey and Gutmann, Sam and Sipser, Michael},
  journal = {arXiv:0001106},
  year = {2000},
  doi = {10.48550/arXiv.quant-ph/0001106},
  url = {https://arxiv.org/abs/quant-ph/0001106}
}

@article{Peruzzo2014,
  title={A variational eigenvalue solver on a photonic quantum processor},
  author={Peruzzo, Alberto and McClean, Jarrod and Shadbolt, Peter and Yung, Man-Hong and Zhou, Xiao-Qi and Love, Peter J and Aspuru-Guzik, Al{\'a}n and O’brien, Jeremy L},
  journal={Nature communications},
  volume={5},
  number={1},
  pages={4213},
  year={2014},
  publisher={Nature Publishing Group UK London}
}

@article{lasserre_2001,
author = {Lasserre, Jean B.},
title = {Global Optimization with Polynomials and the Problem of Moments},
journal = {SIAM Journal on Optimization},
volume = {11},
number = {3},
pages = {796-817},
year = {2001},
doi = {10.1137/S1052623400366802},
URL = {https://doi.org/10.1137/S1052623400366802}
}

@ARTICLE{pyqubo,
  author={Zaman, Mashiyat and Tanahashi, Kotaro and Tanaka, Shu},
  journal={IEEE Transactions on Computers}, 
  title={{PyQUBO: Python Library for Mapping Combinatorial Optimization Problems to QUBO Form}}, 
  year={2022},
  volume={71},
  number={4},
  pages={838-850},
  doi={10.1109/TC.2021.3063618}}

@article{PhysRevLett.69.2863,
  title = {Density matrix formulation for quantum renormalization groups},
  author = {White, Steven R.},
  journal = {Phys. Rev. Lett.},
  volume = {69},
  issue = {19},
  pages = {2863--2866},
  numpages = {0},
  year = {1992},
  month = {Nov},
  publisher = {American Physical Society},
  doi = {10.1103/PhysRevLett.69.2863},
  url = {https://link.aps.org/doi/10.1103/PhysRevLett.69.2863}
}

@article{Stoudenmire_12,
   author = "Stoudenmire, E.M. and White, Steven R.",
   title = "Studying Two-Dimensional Systems with the Density Matrix Renormalization Group", 
   journal= "Annual Review of Condensed Matter Physics",
   year = "2012",
   volume = "3",
   number = "Volume 3, 2012",
   pages = "111-128",
   doi = "https://doi.org/10.1146/annurev-conmatphys-020911-125018",
   url = "https://www.annualreviews.org/content/journals/10.1146/annurev-conmatphys-020911-125018",
   publisher = "Annual Reviews",
   issn = "1947-5462",
   type = "Journal Article",
  }

@article{diamond2016cvxpy,
  author  = {Steven Diamond and Stephen Boyd},
  title   = {{CVXPY}: {A} {P}ython-embedded modeling language for convex optimization},
  journal = {Journal of Machine Learning Research},
  year    = {2016},
  volume  = {17},
  number  = {83},
  pages   = {1--5},
}

@article{agrawal2018rewriting,
  author  = {Agrawal, Akshay and Verschueren, Robin and Diamond, Steven and Boyd, Stephen},
  title   = {A rewriting system for convex optimization problems},
  journal = {Journal of Control and Decision},
  year    = {2018},
  volume  = {5},
  number  = {1},
  pages   = {42--60},
}

@article{itensor,
	title={{The ITensor Software Library for Tensor Network Calculations (C++ version)}},
	author={Matthew Fishman and Steven R. White and E. Miles Stoudenmire},
	journal={SciPost Phys. Codebases},
	pages={4},
	year={2022},
	publisher={SciPost},
	doi={10.21468/SciPostPhysCodeb.4},
	url={https://scipost.org/10.21468/SciPostPhysCodeb.4}
}

@article{odono2016scs,
    author       = {Brendan O'Donoghue and Eric Chu and Neal Parikh and Stephen Boyd},
    title        = {Conic Optimization via Operator Splitting and Homogeneous Self-Dual Embedding},
    journal      = {Journal of Optimization Theory and Applications},
    month        = {June},
    year         = {2016},
    volume       = {169},
    number       = {3},
    pages        = {1042-1068},
    url          = {http://stanford.edu/~boyd/papers/scs.html},
}

@Misc{jansen_block_removal,
      title={Block removal for large language models through constrained binary optimization}, 
      author={David Jansen and Roman Rausch and David Montero and Roman Orus},
      year={2026},
      HowPublished={arXiv:2602.00161},
      Doi={arXiv:2602.00161},
      url={https://doi.org/10.48550/arXiv.2602.00161}, 
}

@misc{exclqa_code,
  note = {\url{https://github.com/erikaltelarrea/Excited-Local-Quantum-Annealing}}
}

\end{document}